\documentclass[sigconf]{acmart}

\copyrightyear{2024}
\acmYear{2024}
\setcopyright{acmlicensed}\acmConference[CCS '24]{Proceedings of the 2024 ACM SIGSAC Conference on Computer and Communications Security}{October 14--18, 2024}{Salt Lake City, UT, USA}
\acmBooktitle{Proceedings of the 2024 ACM SIGSAC Conference on Computer and Communications Security (CCS '24), October 14--18, 2024, Salt Lake City, UT, USA}
\acmDOI{10.1145/3658644.3690211}
\acmISBN{979-8-4007-0636-3/24/10}

\usepackage{tikz}
\usepackage{graphicx}
\usepackage{pifont}
\usepackage{makecell,rotating}
\usepackage[ruled,vlined,linesnumbered]{algorithm2e}
\usepackage{algpseudocode}
\usepackage{upgreek}
\usepackage{color}
 
\usepackage{amsmath,amsthm,amssymb,amsfonts}
\usepackage{epstopdf}
\usepackage{tcolorbox}
\tcbuselibrary{breakable}
\tcbuselibrary{skins}
\usepackage{eucal}
\usepackage{verbatim}
\usepackage{subfigure}
\usepackage{xspace}
\usepackage{setspace}
\usepackage{calrsfs}  
\usepackage{multirow}
\usepackage{balance}
\usepackage{booktabs,color}
\usepackage{tabularx}
\usepackage{xcolor}
\usepackage{bbding,enumitem}
\usepackage{caption}
\usepackage{tikz}
\usepackage{filecontents,  threeparttable}
\usepackage{tabularx}
\newcolumntype{Y}{>{\centering\arraybackslash}X}
\usepackage{booktabs}
\usepackage{xcolor}
\usepackage{utfsym}
\usepackage{geometry}

\newcommand{\first}{\textsf{(i)}\xspace}
\newcommand{\second}{\textsf{(ii)}\xspace}
\newcommand{\third}{\textsf{(iii)}\xspace}
\newcommand{\fourth}{\textsf{(iv)}\xspace}
\newcommand{\fifth}{\textsf{(v)}\xspace}
\newcommand{\ours}{\textsf{Oscar}\xspace}

\newcommand{\zxy}[1]{\textcolor{black}{#1}}
\newcommand{\edit}[1]{\textcolor{black}{#1}}

\newcommand{\eg}{\emph{e.g.,}\xspace}
\newcommand{\ie}{\emph{i.e.,}\xspace}

\begin{document}
%-------------------------------------------------------------------------------

\title{Towards Fine-Grained Webpage Fingerprinting at Scale}

\author{Xiyuan Zhao}
\authornote{Both authors contributed equally to this work.}
\affiliation{
  \institution{INSC \& BNRist, Tsinghua University}
  \city{Beijing}
  \country{China}
}
\email{zhaoxy23@mails.tsinghua.edu.cn}

\author{Xinhao Deng}
\authornotemark[1]
\affiliation{
  \institution{INSC \& BNRist, Tsinghua University}
  \city{Beijing}
  \country{China}
}
\email{dengxh23@mails.tsinghua.edu.cn}

\author{Qi Li}
\affiliation{
  \institution{INSC, Tsinghua University}
  \country{}
}
\affiliation{
  \institution{Zhongguancun Laboratory}
  \city{Beijing}
  \country{China}
}
\email{qli01@tsinghua.edu.cn}

\author{Yunpeng Liu}
\affiliation{
  \institution{INSC, Tsinghua University}
   \city{Beijing}
  \country{China}
}
\email{liuyp20@mails.tsinghua.edu.cn}

\author{Zhuotao Liu}
\affiliation{
  \institution{INSC, Tsinghua University}
  \country{}
}
\affiliation{
  \institution{Zhongguancun Laboratory}
  \city{Beijing}
  \country{China}
}
\email{zhuotaoliu@tsinghua.edu.cn}

\author{Kun Sun}
\affiliation{
  \institution{IST, George Mason University}
  \city{Fairfax}
  \country{USA}
}
\email{ksun3@gmu.edu}
% Fairfax, VA, USA

\author{Ke Xu}
\affiliation{
  \institution{DCST, Tsinghua University}
  \country{}
}
\affiliation{
  \institution{Zhongguancun Laboratory}
   \city{Beijing}
  \country{China}
}
\email{xuke@tsinghua.edu.cn}

\renewcommand{\shortauthors}{Xiyuan Zhao and Xinhao Deng, et al.}

%-------------------------------------------------------------------------------
\begin{abstract}
%-------------------------------------------------------------------------------
Website Fingerprinting (WF) attacks can effectively identify the websites visited by Tor clients via analyzing encrypted traffic patterns.
Existing attacks focus on identifying different websites, but their accuracy dramatically decreases when applied to identify fine-grained webpages, especially when distinguishing among different subpages of the same website.
WebPage Fingerprinting (WPF) attacks face the challenges of highly similar traffic patterns and a much larger scale of webpages. 
Furthermore, clients often visit multiple webpages concurrently, increasing the difficulty of extracting the traffic patterns of each webpage from the obfuscated traffic.
In this paper, we propose \ours, a WPF attack based on multi-label metric learning that identifies different webpages from obfuscated traffic by transforming the feature space.
\ours can extract the subtle differences among various webpages, even those with similar traffic patterns.
In particular, \ours combines proxy-based and sample-based metric learning losses to extract webpage features from obfuscated traffic and identify multiple webpages.
We prototype \ours and evaluate its performance using traffic collected from 1,000 monitored webpages and over 9,000 unmonitored webpages in the real world. \ours demonstrates an 88.6\% improvement in the multi-label metric Recall@5 compared to the state-of-the-art attacks.
\end{abstract}




\maketitle

%-------------------------------------------------------------------------------
\section{Introduction}
%-------------------------------------------------------------------------------
The Onion Routing (Tor) has millions of daily active clients and protects their online privacy through multi-layer encryption with multiple randomly selected relays~\cite{tor}.
However, it is vulnerable to Website Fingerprinting (WF) attacks, which can effectively deanonymize the communication.
WF attacks identify the websites visited by Tor clients through analyzing the unique traffic patterns of the websites, \eg packet sizes, timestamps, and directions. 
Prior WF attacks~\cite{AWF, DF, Tik-Tok, BAPM, tmwf, ARES} develop complex model structures to extract features of various websites from traffic.
However, these attacks focus on identifying websites rather than fine-grained webpages. Since a single website often hosts multiple webpages, accurately identifying fine-grained webpages can provide additional valuable information. 
The performance of existing WF attacks significantly declines when tasked with webpage identification, as models trained with cross-entropy loss struggle to capture the subtle differences in webpage traffic. 

To identify different webpages, a series of fine-grained WF attacks have been studied, \ie WebPage Fingerprinting (WPF) attacks~\cite{shen2019webpage, zhang2019deep, lu2021gap, shen2020fine, shen2021efficient}, which leverage both coarse-grained and fine-grained traffic attributes. 
However, Tor clients often visit multiple webpages consecutively, a common behavior that significantly complicates webpage fingerprinting and makes it challenging for existing attacks to extract the unique patterns of each webpage from the obfuscated traffic under the multi-tab setting~\cite{ARES}.
Furthermore, evaluations of previous WF and WPF attacks generally involve a limited number of webpages. As the scale of webpages expands, these attacks experience degrading performance~\cite{cherubin2022online}.
% and increasing training overhead~\cite{ARES}.

To fully relax the assumption of existing attacks on the browsing behaviors of Tor clients, the goal of this paper is to develop a fine-grained webpage fingerprinting attack that is capable of identifying multiple webpages concurrently visited. 
Generally, there are three main challenges in webpage fingerprinting.
First, webpage traffic patterns share higher similarity~\cite{lu2021gap, wang2020high} due to analogous layouts, increasing the difficulty of distinguishing different webpages in the original feature space. Second, clients typically browse multiple webpages concurrently. This will generate obfuscated traffic from multiple webpages, resulting in irrelevant packets that can interfere with the identification of individual webpages. Additionally, multi-tab traffic patterns exhibit higher diversity due to the various combinations of webpages and the dynamic packet order.
Third, the scale of webpages is approximately 50 times larger than that of websites~\cite{webpage_number}, posing significant challenges to the performance of existing methods that primarily focus on website fingerprinting. 

In this paper, we develop a multi-tab WPF attack framework called \ours, which is capable of identifying webpages based on the subtle differences in their traffic patterns. The design of \ours is based on the key observation that even though webpages within the same website share similar layouts, their contents and resources still exhibit differences, leading to subtle variations in their local traffic patterns. 
\zxy{We utilize metric learning to transform the feature space, clustering traffic of the same webpages and distancing traffic of different webpages to extract the subtle differences among webpage traffic in the new feature space.}

It is challenging to develop the metric-learning-based WPF attack under the multi-tab setting due to the class collapse problem, where the traffic of all webpages clusters to a single point~\cite{levi2021rethinking}.
To address this issue, \ours utilizes multi-label metric learning to construct the WPF attack.
First, it employs two data augmentation mechanisms based on the characteristics of multi-tab traffic to enhance the sample diversity.
Second, it utilizes metric learning to transform the feature space to separate different webpages based on a multi-label loss, which includes a proxy-based loss and a sample-based loss.
Specifically, to address the class collapse issue, \ours clusters relevant webpage traffic by setting proxies for each webpage and separates irrelevant webpage traffic by isolating samples with low label correlation. 
Finally, \ours achieves efficient and accurate multi-tab webpage identification on a large scale of monitored webpages by leveraging proxy-based and sample-based combined k-NN classifiers.
The webpage identification relies on the distribution characteristics in the transformed feature space.

We evaluate \ours using datasets collected under both the closed-world and open-world settings, including 1,000 monitored webpages and more than 9,000 unmonitored webpages. 
To the best of our knowledge, these are the first multi-tab webpage traffic datasets regarding each webpage as a distinct class, with the number of concurrently accessed webpages being dynamic.
We release the datasets and source code of \ours\footnote{https://zenodo.org/records/13383332}.
Compared to state-of-the-art attacks, \ours achieves an average improvement of 88.6\% and 76.7\% in Recall@5 under the closed-world and open-world settings, respectively.
Moreover, \ours significantly outperforms previous attacks under various scales of monitored webpages.

In summary, the contributions of this paper are four-fold:
\begin{itemize}[leftmargin=*]
  \item We develop a fine-grained WPF attack, \ours, to identify webpages from obfuscated traffic under the multi-tab setting. To the best of our knowledge, \ours is the first multi-tab attack to accurately identify fine-grained webpages at scale. 

  \item We utilize data augmentation techniques both within and between traffic samples, leveraging the characteristics of multi-tab traffic. The data augmentation designed for multi-tab traffic enhances the generalization of \ours in real-world WPF attacks.
  
  \item We develop a multi-label metric learning method for traffic samples to transform the feature space. The feature transformation combines the proxy-based loss and sample-based loss, effectively separating different webpages from multi-tab obfuscated traffic.
 
  \item We collect the first multi-tab webpage traffic datasets with 1,000 monitored webpages and over 9,000 unmonitored webpages, and validate the performance of \ours under both the closed-world and open-world settings. 

\end{itemize}
%-------------------------------------------------------------------------------
\section{Background}
%-------------------------------------------------------------------------------

\subsection{WF and WPF Attacks}
\noindent \textbf{WF Attacks.} 
Recently, encrypted traffic analysis has been extensively studied~\cite{qing2023ndss, li2022tifs}.
WF attacks, \zxy{a specific approach within the broader field of encrypted traffic analysis,} aim to identify the unique traffic patterns of websites, including packet time intervals, sizes, and directions. 
WF attacks compromise the privacy of Tor clients by extracting the traffic patterns of different websites from packet sequences.
ML-based WF attacks~\cite{kFP, CUMUL, wang2014effective} utilize expert knowledge to construct features specific to websites for identification. 
With advanced deep learning (DL) algorithms, DL-based WF attacks~\cite{DF, Var-CNN} enable automatic feature extraction and robust attacks. 
To apply WF attacks in the real world, existing works develop WF attacks for various real-world scenarios, such as under the multi-tab setting~\cite{ARES, BAPM, tmwf}, limited training data~\cite{TF, gandalf}, dynamic network conditions~\cite{2023aug}, and various defenses~\cite{Tik-Tok, shen2023subverting}. 
However, existing WF attacks focus on identifying different websites, and have mostly been evaluated on index pages.
Despite certain attacks gathering subpage traffic samples~\cite{ARES, gandalf, goodenough}, their identification targets remain restricted to websites.
\zxy{When applied to fine-grained webpage identification, these attacks become ineffective due to performance deterioration caused by the high similarity of webpage traffic patterns.}

\noindent \textbf{WPF Attacks.} 
Existing ML-based WPF attacks~\cite{shen2019webpage, shen2020fine, zhang2019deep} leverage global and local features to differentiate webpages.
Existing DL-based WPF attacks~\cite{lu2021gap, shen2021efficient} apply powerful Convolutional Neural Networks (CNN) or Graph Neural Networks (GNN) to extract features from webpage traffic.
However, these methods either consider distinguishing webpages within the same website or suffer from the limitation of small-scale webpages.
Additionally, none of the above works address the challenge of multi-tab identification.

\begin{table}
\small
  \centering
  \caption{Summary of Existing Attacks.}
  \label{tab:sota}
  \setlength{\tabcolsep}{4mm}{
  \begin{threeparttable}
  \scalebox{0.87}{
  \begin{tabular}{c|c|c|c}
   \toprule
    \textbf{Attacks}  & \textbf{\makecell[c]{Identification \\ target\tnote{1}}}  & \textbf{Multi-tab\tnote{2}}  & \textbf{\makecell[c]{Large scale\tnote{3}}}\\
    
    \midrule
    k-FP~\cite{kFP} & Website & \usym{2717} & \usym{2717} \\
    DF~\cite{DF}  & Website & \usym{2717} & \usym{2717} \\
    Tik-Tok~\cite{Tik-Tok} & Website & \usym{2717} & \usym{2717} \\
    TF~\cite{TF} & Website & \usym{2717}  & \usym{2717} \\
    MWF~\cite{MWF, MWF2} & Website & \usym{2713}  & \usym{2717} \\
    BAPM~\cite{BAPM} & Website & \usym{2713}  & \usym{2717} \\
    NetCLR~\cite{2023aug} & Website & \usym{2717} &  \usym{2717} \\
    TMWF~\cite{tmwf} & Website & \usym{2713} & \usym{2717} \\
    ARES~\cite{ARES} & Website & \usym{2713} & \usym{2717} \\
    FineWP~\cite{shen2019webpage, shen2020fine} & Webpage & \usym{2717} & \usym{2717} \\
    BurNet~\cite{shen2021efficient} & Webpage & \usym{2717} &  \usym{2717} \\
    GAP-WF~\cite{lu2021gap} & Webpage & \usym{2717}  & \usym{2717} \\
    \ours & Webpage & \usym{2713}  & \usym{2713} \\
    
    \bottomrule
  \end{tabular}}
  
  \begin{tablenotes}    
        \footnotesize              
        \item[1] \textbf{Identification target} refers to if the attacks target on websites or webpages.
        \item[2] \textbf{Multi-tab} refers to whether the attacks consider the setting where clients can visit \\ multiple webpages and traffic from various webpages mixes.
        % \item[3] \textbf{Generalization} refers to whether the attacks consider identifying webpages from \\ the traffic of unknown webpage combinations under the multi-tab setting.
        \item[3] \textbf{Large scale} refers to whether the attacks are evaluated under the \\ datasets with thousands of monitored webpages.
        
  \end{tablenotes}
  \end{threeparttable}}
\end{table}

The existing attacks are summarized in Table~\ref{tab:sota}.
Different from existing attacks, \ours implements multi-tab identification targeting specific webpages from various websites.
Furthermore, \ours expands the scale of monitored webpages to realize the real-world WPF attack.

\subsection{Metric Learning}
Metric learning aims to develop a new data representation to establish the similarity of samples, enhancing the ability to distinguish among samples from different classes~\cite{metric-learning3}.
This technique is widely applied in the image recognition domain to improve the prediction accuracy in both classification and clustering problems~\cite{hadsell2006dimensionality,triplet-loss,n-pair-loss,histogram-loss,structured-loss,angular-loss,hierarchical-triplet-loss,multi-similarity-loss,proxy-nca,kim2020proxy}.
The core concept of metric learning involves comparing anchor samples with positive samples or/and negative samples during the learning process, thereby pulling samples from the same class closer together and pushing samples from different classes farther apart.

TF~\cite{TF} and NetCLR~\cite{2023aug} utilize metric learning and contrastive learning to identify the differences among website traffic. Metric learning is usually based on labeled data, while contrastive learning does not require samples to be labeled.
However, these attacks depend on clean traffic under the single-tab setting to transform the feature space. In particular, the performance of TF is significantly impacted by the class collapse issue~\cite{levi2021rethinking} when the traffic is obfuscated under the multi-tab setting.
Moreover, NetCLR augments traces by analyzing single-tab traffic to pre-train a model using self-supervised learning. However, there is a significant gap between single-tab and multi-tab traffic patterns.
% Moreover, NetCLR augments traces based on the analysis of single-tab traffic to pre-train a model based on self-supervised learning, whereas a significant gap exists in single-tab and multi-tab traffic patterns.
Therefore, we design a multi-label metric learning method to distinguish webpages from obfuscated traffic.

%-------------------------------------------------------------------------------
\section{Threat Model} %and Assumption}
%-------------------------------------------------------------------------------

The threat model of \ours is illustrated in Figure 1. In our threat model, the clients are free to explore various webpages, moving beyond just the index pages. They are redirected to other webpages by clicking on links displayed on the current webpage. Clients are often interested in multiple webpages, therefore they may open multiple webpages concurrently, and the number of webpages is dynamic and unknown to the attacker a prior. Furthermore, a website typically hosts multiple webpages. Therefore, the scale of webpage monitoring is far larger than that of website monitoring. We broaden the scope of our monitoring to include more monitored webpages, accommodating the size of real-world webpages.

To identify the webpages visited by the client, the attacker has the ability to eavesdrop on the communication link between the client and the guard node of the Tor network to analyze the patterns of bidirectional communication packets~\cite{DF, ARES, tmwf, TF}. Though the connections between the client and the guard node are protected by encryption, the attacker can gather traffic metadata such as the direction, size, and time sequences of traffic packets to extract unique fingerprint features that can be used to identify the webpages visited by the client. 
In addition, the attacker does not require the ability to actively modify, delay, or decrypt these packets.

Similar to existing WF attacks~\cite{AWF, DF, ARES, Tik-Tok, Var-CNN, tmwf}, we consider both the closed-world and open-world settings. Under the closed-world setting, Tor clients can only browse a limited number of webpages, \ie the monitored webpages. The attacker can collect traffic from all monitored webpages to train the model.
Under the open-world setting, Tor clients can freely access a large number of webpages unknown to the attacker, \ie the unmonitored webpages.
Given the vast number of webpages in reality, obtaining training samples for all of them is impractical, highlighting the realism of open-world identification.
Note that Tor clients under the open-world setting still use multiple tabs to access both the monitored and unmonitored webpages.

\begin{figure}
  \centering
  \includegraphics[width=0.85\linewidth]{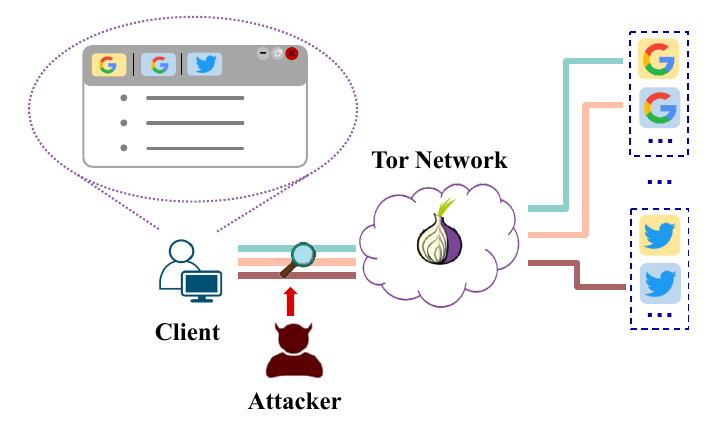}
  \caption{The threat model of \ours, where clients can access both the index pages and other webpages under the multi-tab setting.}
  \label{fig:threat model}
\end{figure}

%-------------------------------------------------------------------------------
\section{System Overview}
%-------------------------------------------------------------------------------

\begin{figure}
	\begin{minipage}[t]{0.49\linewidth}
		\centering
		\includegraphics[width=1.6in]{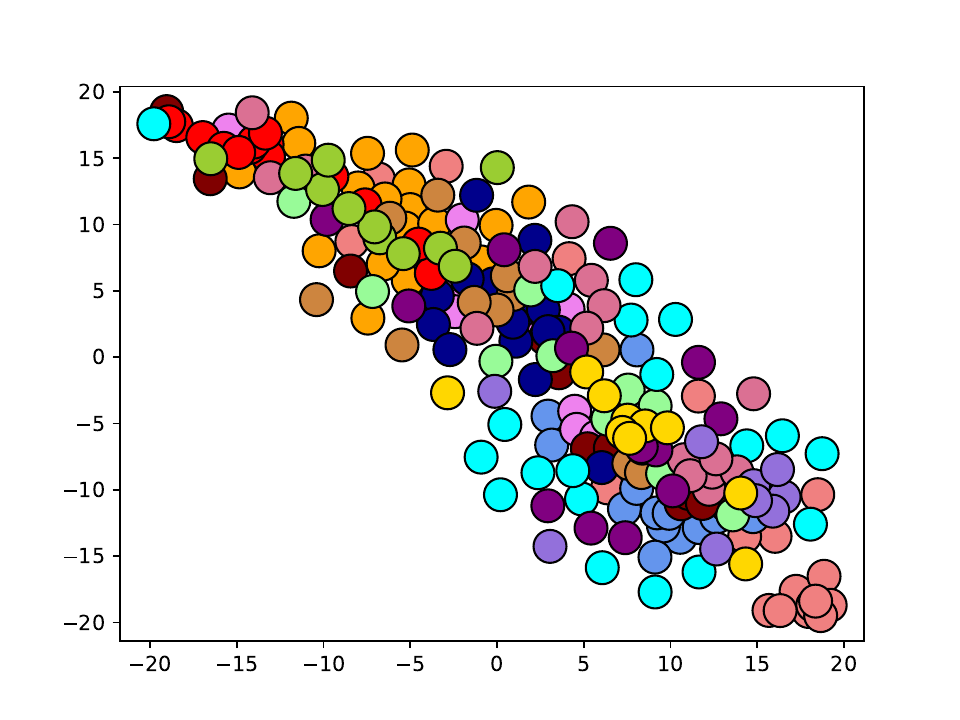}
	\end{minipage}
	\begin{minipage}[t]{0.48\linewidth}
		\centering
		\includegraphics[width=1.6in]{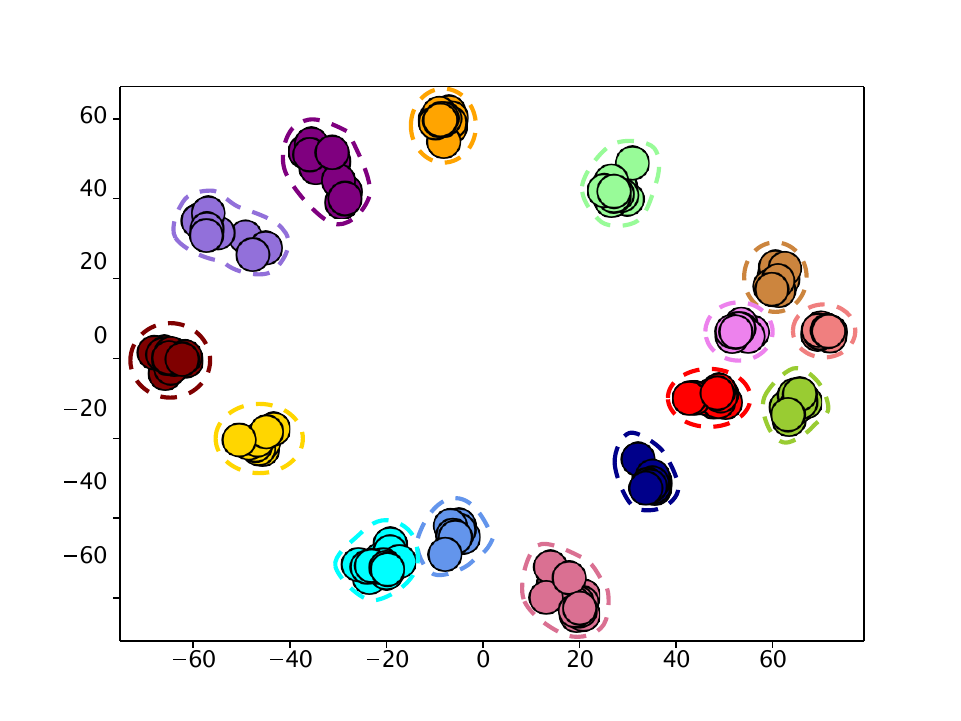}
	\end{minipage}
    \begin{minipage}[t]{1.0\linewidth}
		\centering
		\includegraphics[width=3.1in]{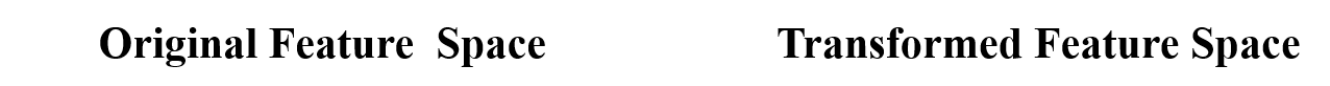}
	\end{minipage}
	\caption{Feature transformation based on multi-label metric learning. Each color represents samples from a distinct webpage.}
	\label{fig:intuition}
\end{figure}

\ours is a robust WPF attack that leverages the differences in webpage traffic to identify the webpages visited by Tor clients. 
Since different webpages from the same website are not identical, it leads to variations in local traffic patterns. Although it is difficult to directly distinguish webpage traffic in the original feature space, by extracting the differences in webpage traffic through metric learning, we can separate different webpages in the transformed feature space. 
Figure~\ref{fig:intuition} shows the feature transformation of webpage traffic through metric learning, where traffic from different webpages is isolated.
\zxy{Specifically, in our study of 1,000 webpages, the similarity of traffic features between different webpages decreases by an average of 52.92\% after feature transformation.
This reduction in similarity highlights the effectiveness of \ours in distinguishing webpage traffic and accurately identifying the visited webpages.}

\begin{figure*}
  \centering
  \includegraphics[width=1\linewidth]{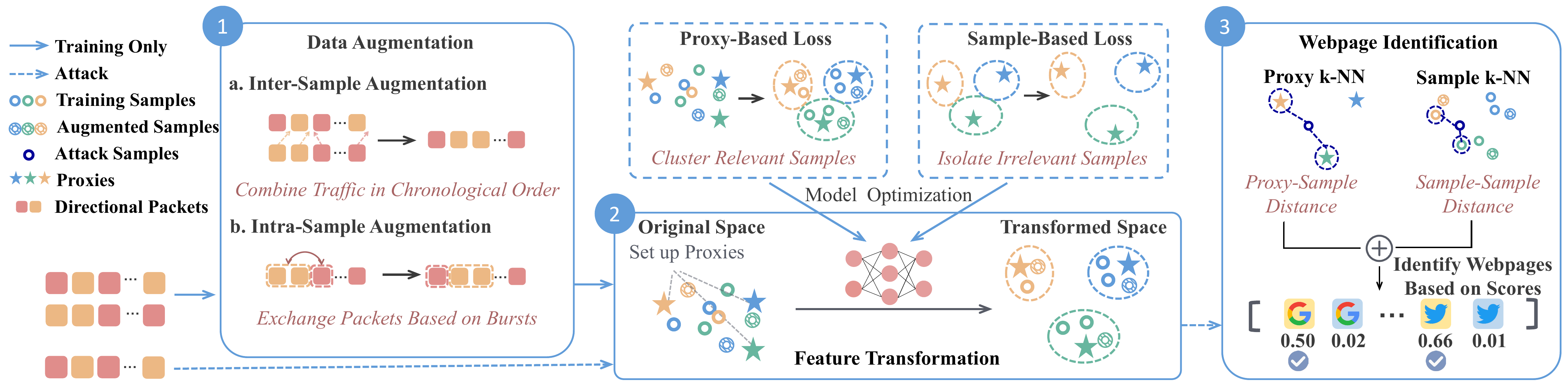
  }
  \caption{Design overview of \ours, which consists of three modules: Data Augmentation module, Feature Transformation module and Webpage Identification module.}
  \label{fig:design overview}
\end{figure*}

Formally, considering W monitored webpages, the problem is defined as follows: the sample set X=\{$x_1$,..., $x_N$\} contains N samples, where each sample holds the dimension of $d_i*2$, comprising both the direction sequences $ds_i$ and time sequences $ts_i$ extracted from the original packet sequences. The direction sequences are distinguished by +1 and -1 for outgoing and incoming packets, and the time sequences are calculated as the interval relative to the first packet.
The label set is defined as Y=\{$y_1$,... , $y_N$\}, where each label {$y_i$} is a W-dimensional 0/1 vector. $y_ {ij} = 1$ shows that the ith sample comprises the traffic from the jth webpage. Under our attack setting, the clients can visit multiple webpages, \ie $|y_i| \geq 1$. Consequently, our problem is characterized as a multi-label classification, where each sample may be associated with one or more labels. The objective is to precisely determine if each label is present in samples.
This is more challenging than the multi-class classification, as the latter only considers the label with the highest probability.

Figure~\ref{fig:design overview} shows the three modules in \ours to achieve the robust WPF attack.
First, the Data Augmentation module enhances the generalization of \ours through inter-sample and intra-sample combined data augmentation operations based on the characteristics of multi-tab traffic in the real world.
\edit{The inter-sample augmentation combines the traffic of two samples in chronological order to enhance the traffic diversity from different webpage combinations. The intra-sample augmentation exchanges packets within a single sample to accommodate variations in packet order.}
Second, the Feature Transformation module leverages both the original samples and augmented samples to train the DF-based feature transformation model based on two losses. The proxy-based loss clusters relevant samples, and the sample-based loss isolates irrelevant samples. Combining these two losses, this module obtains a feature transformation model responsible for transforming the feature space.
Third, the Webpage Identification module achieves efficient multi-label webpage classification \zxy{based on the distribution features of webpage traffic in the transformed feature space}. It is composed of two k-NN classifiers, incorporating both the proxy-sample distance and the sample-sample distance, to enhance accurate webpage identification. The predicted webpages are determined based on the scores calculated from the two classifiers.

\section{System Design}

In this section, we present the design details of \ours, including the Data Augmentation module, the Feature Transformation module, and the Webpage Identification module.

\subsection{Data Augmentation}
The Data Augmentation module employs inter-sample and intra-sample operations to generate simulated traffic based on the original traffic, thereby enhancing the generalization to diverse multi-tab traffic.
To achieve the data augmentation, \ours operates on the raw traffic samples and labels based on traffic characteristics.
\edit{However, clients typically browse multiple webpages concurrently.
Existing data augmentation operations~\cite{tripod, 2023aug, HDA} cannot be applied to multi-tab traffic}, as they ignore the diversity introduced by traffic mixing under the multi-tab setting. 
To address this issue, we design two data augmentation operations specifically for multi-tab traffic.

\begin{figure}[t!]
\centering
    \begin{minipage}[t]{1\linewidth}
		\centering
		\includegraphics[width=1.7in]{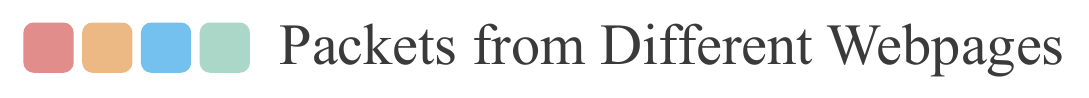}
	\end{minipage}
  \subfigure[Inter-Sample Augmentation]{
        \label{fig:mixing}
        \includegraphics[width=0.49\linewidth]{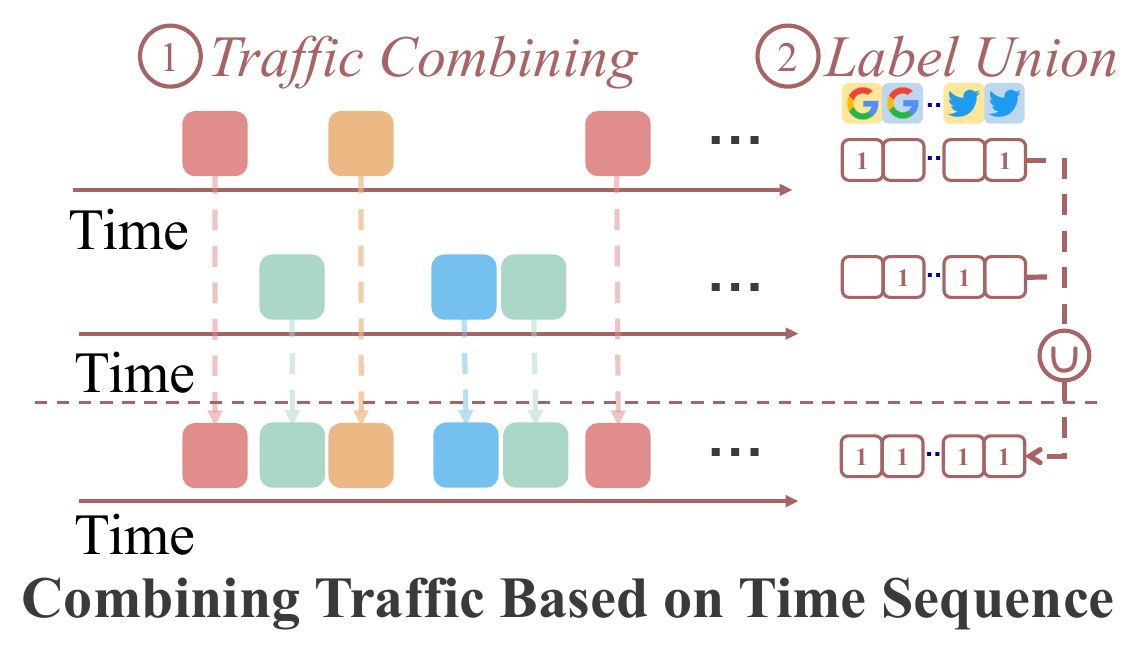}
    }
    \subfigure[Intra-Sample Augmentation]{
        \label{fig:exchanging}
        \includegraphics[width=0.455\linewidth]{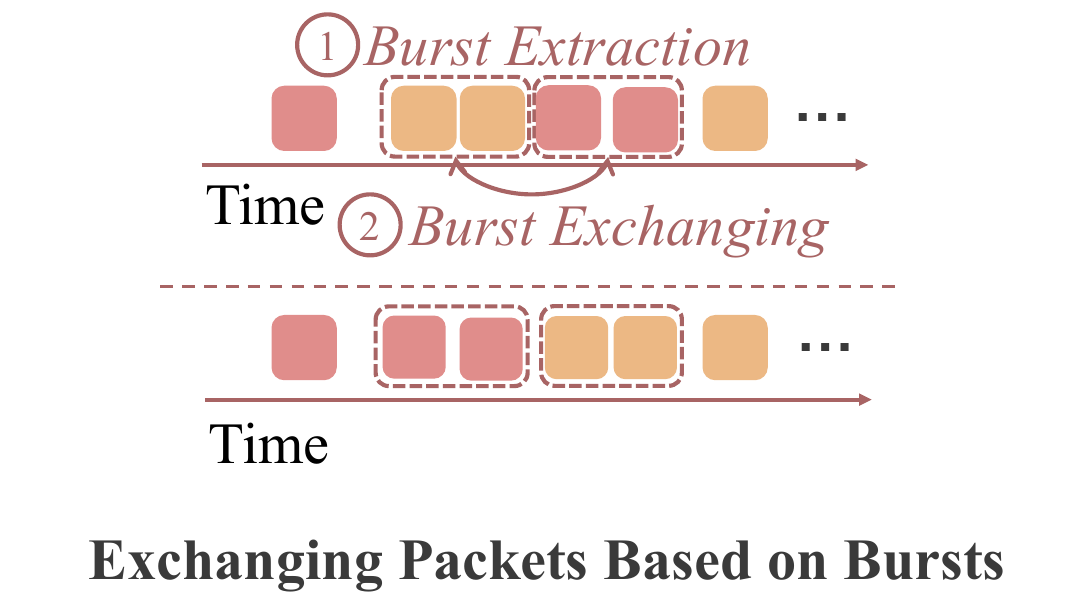}
    }
  \caption{Two operations in the Data Augmentation module: inter-sample and intra-sample augmentation. Packets from different webpages are shown in different colors.}
	\label{fig:augmentation operation}
\end{figure}

\edit{Multi-tab traffic exhibits greater diversity compared to single-tab traffic as the clients may browse multi-tab webpages with different webpage combinations, resulting in various mixed traffic. 
Even when the same combination of webpages is accessed, packets from different webpages within a session are transmitted through different circuits, leading to a dynamic packet order~\cite{packet-disordering}.
Therefore, multi-tab traffic is more diverse, both for different webpage combinations and the same webpage combinations.
To enrich the sample diversity under the multi-tab setting, we incorporate two data augmentation operations, as shown in Figure~\ref{fig:augmentation operation}.
First, we design an inter-sample data augmentation that combines traffic of different samples to adapt to the diversity of various webpage combinations.
Second, we implement an intra-sample data augmentation that exchanges packets within a single sample to handle the variability in packet order, addressing the diversity of traffic within the same webpage combinations.
% Second, we design an intra-sample data augmentation that exchanges packets within a single sample to accommodate the diversity of the same webpage combinations caused by variations in packet order.
We detail these two operations as follows.}

\begin{figure}[htbp]
    \flushleft
    \subfigure[Edit distance of samples]{
        \label{fig:edit_distance}
        \includegraphics[width=0.44\linewidth]{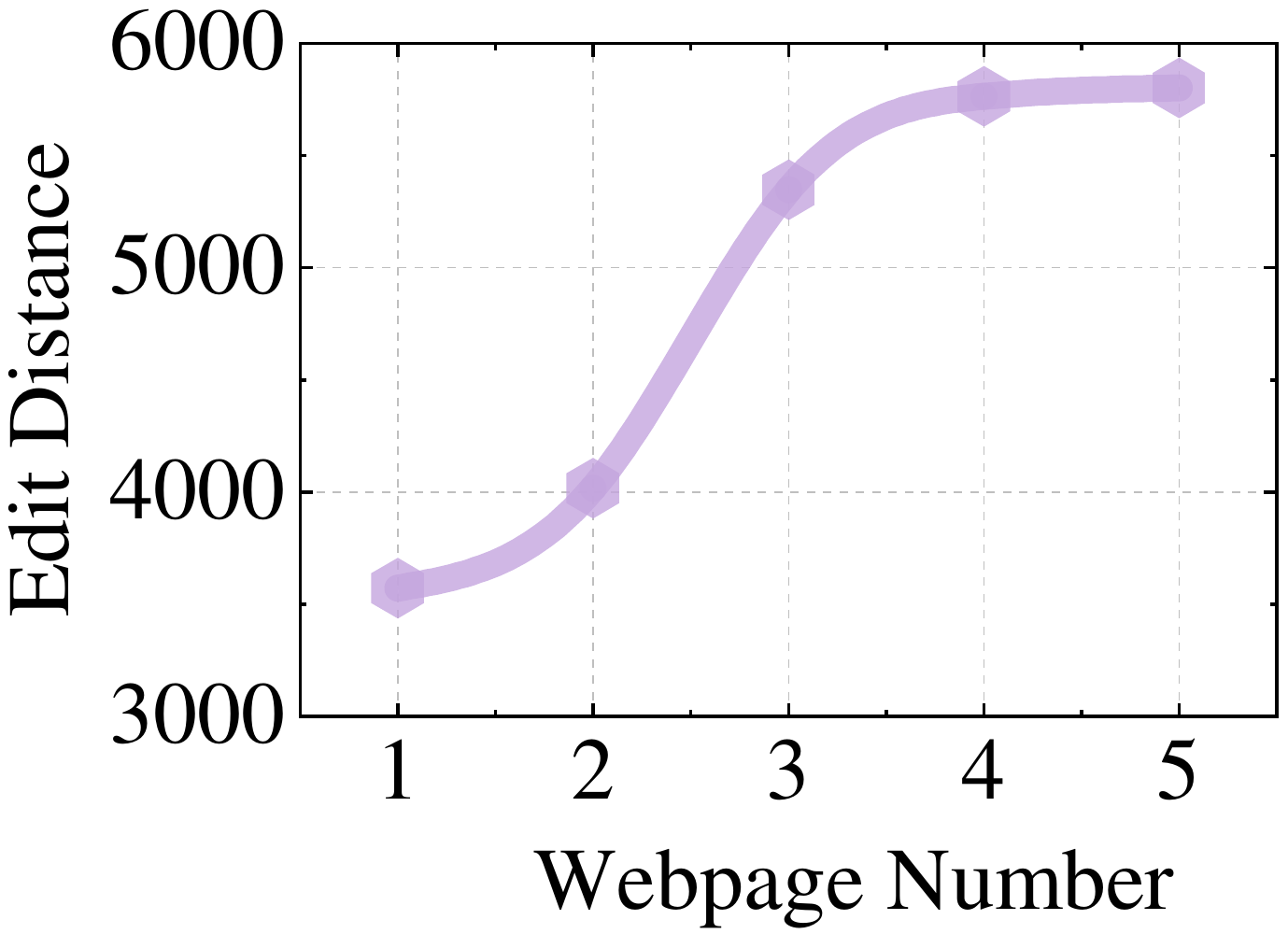}
    }
    \subfigure[Burst patterns when browsing the same webpages twice]{
        \label{fig:sample_contrasting}
        \includegraphics[width=0.43\linewidth]{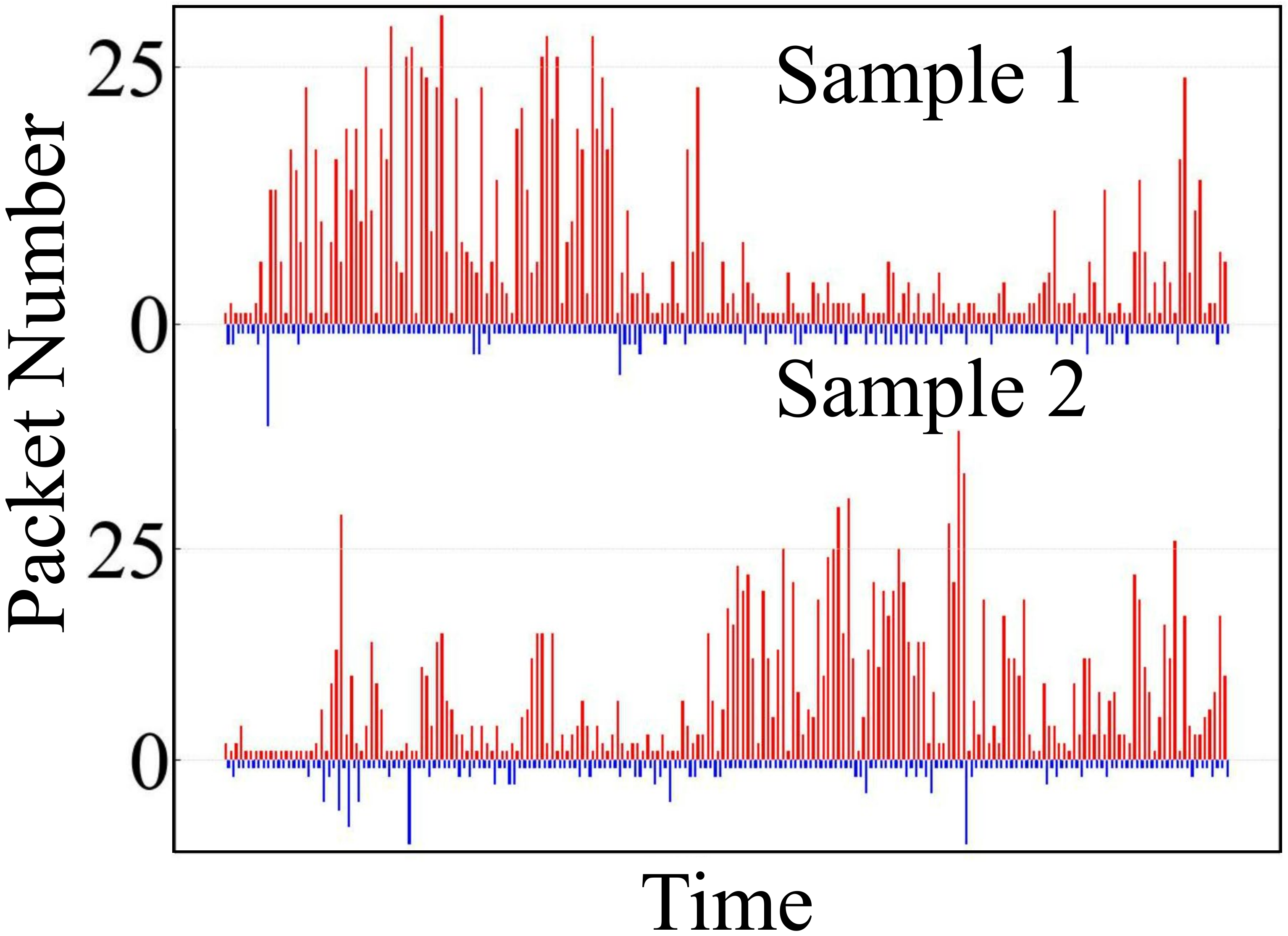}
    }
    
	\caption{Analysis of multi-tab obfuscated traffic collected from the real world.}
	\label{fig:traffic}
\end{figure}

\noindent \textbf{Inter-Sample Augmentation.}
We design an inter-sample data augmentation based on traffic combining to enhance the variety of traffic from different webpage combinations.
Under the multi-tab setting, traffic from various webpages is mixed, making traffic from the same webpage display completely different patterns when mixed with different webpages.
Figure~\ref{fig:edit_distance} illustrates the correlation between the edit distance of samples and the quantity of browsed webpages, where the edit distance reflects the similarity of two vectors by measuring the minimum operations required to transform one vector to another.
Specifically, we select 500 samples from different webpage combinations. These samples are grouped based on the number of labels, ranging from 1 to 5, with 100 samples in each group. We then calculate the edit distance among samples within each group, where all samples share the same number of webpage labels.
The results show that disparity in samples from different webpage combinations increases as the number of concurrently accessed webpages rises.
Therefore, \ours adopts an inter-sample augmentation to adapt to the diversity of various webpage combinations.

Different from the intra-sample augmentation, the inter-sample augmentation involves two samples and operates on both the traffic samples and labels.
Specifically, \ours combines the original two traffic based on the time sequence of packets.
Algorithm~\ref{alg:mixing} details the inter-sample augmentation.
\ours first sets two pointers to indicate the packet indexes of the two samples respectively (line 1).
\ours then compares the times of the indexed packets in the two samples, and adds the packet with the earlier time to the newly generated packet sequence, with the index moving one step backward (line 2-8).
In this way, \ours integrates the packets of the two samples in a time-ordered manner.
To generate the new label, \ours unions the labels of the two samples since the newly generated traffic includes the packets of both the original samples (line 9).
Following this operation, \ours generates a new sample that contains traffic from a webpage combination that is not present during training.
With the inter-sample augmentation, \ours enriches the variety of webpage combinations, enhancing the generalization of \ours when applied to multi-tab webpage identification.

\begin{algorithm}[t]\small
    \SetKwInOut{Input}{input}
    \SetKwInOut{Output}{output} 
    \SetKwComment{Comment}{$\triangleright$\ }{\ }
    \newcommand{\mycommfont}[1]{\scriptsize\textcolor{gray}{#1}}
    \SetCommentSty{mycommfont}
    
    \Input{original sample1: $ds_i$(direction sequence),    \\
           \hspace{2cm} $ts_i$(time sequence);\\
           original label1: $y_i$; \\
          original sample2: $ds_j$(direction sequence),
          \\ \hspace{2cm} $ts_j$(time sequence);\\
          original label2: $y_j$;\\
          input dimension: $d_i$;}
    \Output{generated sample: $ds_g$(direction sequence);\\
           generated label: $y_g$;}
    \BlankLine 

    $index_1 \leftarrow 0$  \hspace{0.3cm} $index_2 \leftarrow 0$
    \tcp*[f]{set two pointers} \\
    \For{$k\leftarrow 0$  \KwTo $d_i$}{
        \eIf(\tcp*[f]{compare the times of the two indexed packets}){$ts_i[index_1] \leq ts_j[index_2]$}{
            % \tcp*[f]{inter-arrival time below $t_{int}$ marks a latent burst} \\
            $ds_g[k] \leftarrow ds_i[index_1]$
            \tcp*[f]{add packet to the new sequence}\\
            $index_1 \leftarrow index_1 + 1$
        }
        {
            $ds_g[k] \leftarrow ds_j[index_2]$
            \tcp*[f]{add packet to the new sequence}\\
            $index_2 \leftarrow index_2 + 1$
        }
    }
    $y_g \leftarrow y_i \cup y_j$ 
    \tcp*[f]{union the labels of two original samples}\\
    return $ds_g$, $y_g$
    \caption{Inter-Sample Data Augmentation.}
    \label{alg:mixing} 
\end{algorithm}

\noindent \textbf{Intra-Sample Augmentation.}
In addition to the inter-sample augmentation, we design an intra-sample data augmentation to adapt to the traffic diversity of the same webpage combinations under the multi-tab setting.
Web browsing on various tabs is based on distinct Tor circuits, resulting in dynamic packet ordering when traffic mixes~\cite{packet-disordering}.
Figure~\ref{fig:sample_contrasting} demonstrates the burst patterns of traffic when browsing the same webpages twice.
It is evident that there is a distinct discrepancy between the burst patterns of the two samples.
To accommodate this sequential diversity when multiple webpages are loaded concurrently, \ours employs an intra-sample augmentation strategy based on packet exchanging.

The intra-sample augmentation modifies traffic within a single sample without altering its label.
Specifically, \ours adopts an exchanging operation based on bursts.
Bursts are consecutive packets of the same direction, often containing resources like texts and images~\cite{2023aug}.
Algorithm~\ref{alg:exchanging} details the exchanging operation.
\ours begins with identifying the bursts in the original direction sequence (line 1).
\ours then determines the number of bursts to be exchanged $ex\_num$ based on the total burst number of the sample with a ratio of $m_e$, and samples $ex\_num$ bursts in the original traffic (line 3-4).
Finally, for each selected burst, \ours exchanges it with the following burst, while leaving the remaining bursts unchanged (line 5-6).
By setting the exchanging ratio advisedly, \ours dynamically adjusts the number of exchanges according to the total burst number of different samples, thus ensuring that the modification of each sample is controlled within a manageable range.
In summary, this operation improves the ability to cope with the dynamic and unpredictable patterns of packet order in multi-tab traffic.

Notably, our data augmentation is grounded in the analysis of multi-tab traffic characteristics. 
By combining inter-sample and intra-sample augmentation techniques, \ours significantly enhances sample diversity, ensuring the generalization of the WPF attack under the multi-tab setting.
After generating augmented samples, we blend them with the original samples for feature transformation.

\begin{algorithm}[t]\small
    \SetKwInOut{Input}{input}
    \SetKwInOut{Output}{output} 
    \SetKwComment{Comment}{$\triangleright$\ }{\ }
    \newcommand{\mycommfont}[1]{\scriptsize\textcolor{gray}{#1}}
    \SetCommentSty{mycommfont}
    
    \Input{original sample: $ds_i$(direction sequence);\\
           exchanging ratio: $m_e$;} 
    \Output{generated sample: $ds_g$(direction sequence);} 
    \BlankLine 

    $burst\_sequences \leftarrow extract\_bursts(ds_i) $ \\
    $ds_g \leftarrow ds_i$ \\
    $ex\_num \leftarrow len(burst\_sequences) * m_e $
    \tcp*[f]{calculate the number of bursts to be exchanged based on the total burst number}\\
    $ex\_bursts \leftarrow sample(burst\_sequences, ex\_num) $
    \tcp*[f]{sample $ex\_num$ bursts from the burst sequence}\\
    \For{$burst\ in\ ex\_bursts$}{
        $exchange\_bursts(ds_g[burst],ds_g[burst+1])$ \tcp*[f]{exchange the selected bursts with their subsequent bursts}
    }
    return $ds_g$
    \caption{Intra-Sample Data Augmentation.}
    \label{alg:exchanging} 
\end{algorithm}

\subsection{Feature Transformation}
The Feature Transformation module transforms traffic features to cluster traffic of the same webpages and separate traffic from different webpages. 
To realize the above feature transformation, \ours needs to contrast the traffic of different webpages based on metric learning, so as to extract the subtle differences in the traffic patterns of different webpages.
However, existing website fingerprinting attacks based on metric learning~\cite{TF} cannot be applied to webpage identification.
The reason is that Tor clients usually browse multiple webpages under the multi-tab setting, with each traffic having multiple labels. 
Traditional metric learning methods select positive and negative samples based on the single label, leading to a dramatic increase in the number of positive samples under the multi-tab setting, which causes the class collapse issue~\cite{levi2021rethinking} (\ie traffic from all different webpages clusters together in the new feature space).

\begin{figure}[t]
  \includegraphics[width=0.9\linewidth]{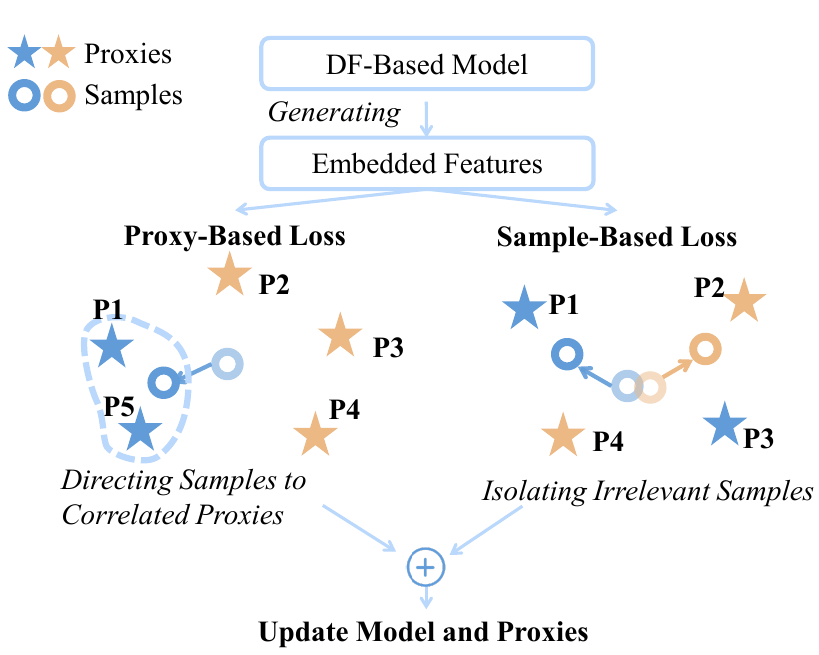}
  \caption{Details of the Feature Transformation module, which comprises a DF-based model and a multi-label metric learning loss function. We display the samples and their correlated proxies in the same color.}
  \label{fig:feature_transformation}
\end{figure}

To address the above issue, we design a multi-label metric learning method to achieve feature transformation. 
The details of the feature transformation are shown in Figure~\ref{fig:feature_transformation}. To effectively identify webpages corresponding to multi-tab obfuscated traffic, our feature transformation contains a feature transformation model to embed features to a lower-dimensional feature space and a multi-label metric learning loss function to aggregate traffic of the same webpages and separate traffic of different webpages in the transformed feature space.

The feature transformation model takes the original direction sequences of traffic as input, and outputs the transformed lower-dimensional vectors.
We select DF as the feature transformation model for the following reasons: \first DF has demonstrated effectiveness in WF attacks, achieving 98\% accuracy in identifying different websites~\cite{DF};
\second DF is built upon CNN, which can effectively extract the features regardless of the part in which the feature fragments appear.
The shift-invariance characteristic of CNN can extract specific features with dynamic locations, which is particularly important under the multi-tab setting.
DF contains four basic convolutional blocks and two fully connected layers.
Each block contains two one-dimensional convolutional layers and one max pooling layer.
We retain the original four convolutional blocks and replace the fully connected layers with a linear layer to embed the features to a low-dimensional feature space, enabling DF to function as a feature extractor.

Beyond the feature transformation model, the loss function is crucial for the effectiveness of feature transformation.
\ours utilizes a multi-label metric learning loss to aggregate traffic from the same webpages and separate traffic from different webpages.
The loss function comprises two parts: a proxy-based loss and a sample-based loss.

\noindent\textbf{Proxy-Based Loss.} 
As discussed above, existing metric learning approaches calculate loss based on positive samples, which leads to class collapse under the multi-tab setting. 
Therefore, we develop a proxy-based loss to aggregate traffic from the same webpages under the multi-tab setting.
The left part of Figure~\ref{fig:feature_transformation} illustrates the proxy-based loss.
\zxy{We first set up proxies as representatives for each webpage.
Instead of pulling samples with the same labels closer together, the proxy-based loss directs samples to the correlated proxies, therefore effectively aggregating samples of the same webpages.}
Specifically, the positions of the proxies are dynamic and optimized together with the model's parameters in each epoch of model training.
Such adaptability of proxies is crucial as it contributes to learning more accurate distributions of different webpages along with the model progressing through training.

To calculate the proxy-based loss, we first initialize the W proxies in the proxy set P = $\{p_1,...,p_W\}$, where W is the number of webpages.
Proxies hold the dimension of $d_o$, consistent with the embedded vector dimension after the feature transformation model.
After proxy initialization, we excavate the proxy-sample relationship to cluster relevant samples. 

The proxy-sample relationship can be divided into two types: positive proxy-sample pair and negative proxy-sample pair.
In the case where the sample contains traffic from the webpage associated with the proxy, they are identified as a positive proxy-sample pair, otherwise, they are a negative proxy-sample pair, \ie if $y_{ij}=1$, sample $x_i$ and proxy $p_j$ constitute a positive proxy-sample pair.
For positive proxy-sample pairs, the anticipated similarity between them should approach 1, indicating that the samples are close to the relevant proxies in the feature space.
The positive loss is then defined as the difference between the cosine similarity and 1:
\begin{equation}
\begin{aligned}
L_{pos\_proxy}\langle x_i,p_j\rangle=1-cos\_sim\langle x_i,p_j\rangle,
\end{aligned}
\end{equation}
where $cos\_sim$ refers to the cosine similarity shown as follows:
\begin{equation}
\begin{aligned}
cos\_sim\langle x_i,p_j\rangle = \frac{x_i\cdot p_j}{\|x_i\| \times\|p_j\|}.
\end{aligned}
\end{equation}

On the other hand, negative proxy-sample pairs are anticipated to exhibit low similarity, indicating their separation in the transformed feature space.
To prevent overfitting, \ours sets a margin for the expected similarity. If the similarity is below this margin, indicating effective separation from unrelated proxies, the loss is set to zero.
The negative loss is then defined as the maximum of the two terms:
% \vspace{-3pt}
\begin{equation}
\begin{aligned}
L_{neg\_proxy}\langle x_i,p_j\rangle=max(cos\_sim\langle x_i,p_j\rangle-margin,0),
\end{aligned}
\end{equation}
where the margin is a hyperparameter preset. 

We calculate the sum of all positive proxy-sample loss $L_{all\_pos\_proxy}$ and negative proxy-sample loss $L_{all\_neg\_proxy}$, and combine them for the total proxy-based loss $L_{proxy}$:
\vspace{-3pt}
\begin{equation}
\begin{aligned}
L_{proxy}\ =\frac{L_{all\_pos\_proxy}}{\Theta_{pos\_proxy}} + \frac{L_{all\_neg\_proxy}}{\Theta_{neg\_proxy}},
\end{aligned}
\end{equation}
where $\Theta_{pos\_proxy}$ and $\Theta_{neg\_proxy}$ refer to the total number of positive proxy-sample pairs and negative proxy-sample pairs. Note that we divide by the number of positive pairs and negative pairs to avoid quantity imbalance.
The multi-label proxy loss effectively enhances the accuracy of webpage identification by ensuring that samples are tightly distributed around the relevant proxies.
As a result, the patterns of different webpages can be extracted.

\noindent\textbf{Sample-Based Loss.}
Different from the proxy-based loss, \ours utilizes a sample-based loss to separate irrelevant webpage traffic in the transformed feature space. 
Specifically, under the multi-tab setting, the proxy-based loss might inadvertently bring traffic from unrelated webpages closer together. 
The right part of Figure~\ref{fig:feature_transformation} illustrates this conflict.
The left sample, comprising traffic from webpage 1 and webpage 3, is expected to be approximately positioned between proxy 1 and proxy 3. Similarly, the right sample, containing traffic from webpage 2 and webpage 4, is likely to be located between proxy 2 and proxy 4. Despite that these two samples share no common labels, their positions in the transformed feature space can be notably close. 
Therefore, we design a sample-based loss to effectively identify and separate irrelevant webpage traffic by evaluating the similarity between samples in the feature space, ensuring that traffic from distinct webpages is separated.

The sample-based loss takes advantage of the relationship between samples.
Due to large-scale monitored webpages, the number of sample pairs with different labels is very large. Therefore, we design sample mining based on the label coincidence degree to selectively separate sample pairs with low correlation, \ie samples without identical labels.
To mine irrelevant sample pairs, we first sift out the samples with at least two labels and build a new set $X'=\{x_i|x_i \in X \ \land \  |y_i|>1\}$, where the total number of the filtered samples is N'. Then we find samples with no overlapping labels in X', \ie $y_i*y_j=0$, and form the sample pairs. For irrelevant sample pairs, the cosine similarity between them is defined as:
\begin{equation}
\begin{aligned}
cos\_sim\langle x_i,x_j\rangle = \frac{x_i\cdot x_j}{\|x_i\| \times\|x_j\|}.
\end{aligned}
\end{equation}
\zxy{We expect irrelevant sample pairs to show a lower degree of cosine similarity.}
Similar to the calculation of the negative proxy-sample loss, the margin is applied, and when the similarity is below the margin, the loss is set to 0:
\begin{equation}
\begin{aligned}
L_{ir\_sample}\langle x_i,x_j\rangle=max(cos\_sim\langle x_i,x_j\rangle-margin,0).
\end{aligned}
\end{equation}
In each epoch, we collect all the irrelevant sample pairs in the batch and compute their sum $L_{all\_ir\_sample}$. The total sample loss is calculated as follows:
\begin{equation}
\begin{aligned}
L_{sample}\ =\frac{L_{all\_ir\_sample}}{\Theta_{ir\_sample}},
\end{aligned}
\end{equation}
where $\Theta_{ir\_sample}$ refers to the total number of irrelevant sample pairs.
Compared with the vanilla sample-based metric learning methods, our approach focuses on samples with low label correlation, thus significantly reducing the pair number.
Overall, the sample-based loss enhances the effectiveness of the feature transformation by isolating irrelevant samples.

After calculating the two parts of losses, we combine them for the total loss:
\begin{equation}
\begin{aligned}
Loss = L_{proxy}+\beta\times L_{sample},
\end{aligned}
\end{equation}
where $\beta$ is a hyperparameter that adjusts the weights of these two losses. When $\beta=0$, the above loss simplifies to the multi-label proxy-based loss. 

Note that the proxies are added to the parameters and updated with the model's parameters using the same optimizer.
This dynamic adjustment refines the distributions of proxies, leading to more precise boundaries for each webpage.
Following the feature transformation module, we strategically transform the feature space to distinctly separate different webpages.
This separation is vital for effective webpage classification, as it ensures that each webpage is represented in a unique and distinguishable manner within the feature space.

\subsection{Webpage Identification}
\begin{figure}
  \centering
  \includegraphics[width=0.9\linewidth]{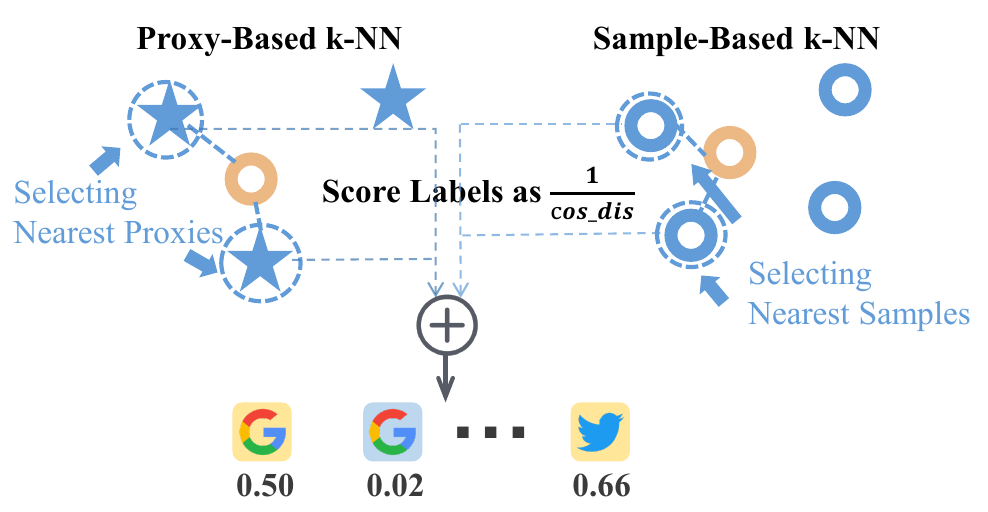}
  \caption{Details of the Webpage Identification module, which consists of two k-NNs to calculate label scores.}
  \label{fig:knn}
\end{figure}

The Webpage Identification module integrates two k-NN classifiers to achieve robust multi-tab webpage identification. 
Traditional k-NN classifiers typically rely on the labels of the nearest samples for classification.
However, the diversity of multi-tab traffic can lead to sample drift and performance degradation. 
Therefore, we integrate a proxy-based k-NN and a sample-based k-NN, as illustrated in Figure~\ref{fig:knn}, to achieve robust webpage identification. 
The proxy-based k-NN benefits from constantly updated and precise representations of webpages, focusing on the uniform features of webpages. Meanwhile, the sample-based k-NN accounts for the diversity of multi-tab webpage traffic across various webpage combinations. 
Finally, we combine the results of these two classifiers to calculate the label scores, leveraging the strengths of both to improve the robustness of webpage identification under the multi-tab setting.

The proxy-based k-NN achieves classification based on the proxy-sample distance. Specifically, it retrieves the nearest $b$ proxies and calculates the scores using the distances of the retrieved proxies with the target sample. Given that our transformed feature space is built upon cosine similarity, our k-NN classifiers adopt the cosine distance for score calculation:
\begin{equation}
\begin{aligned}
cos\_dis\langle x_{target},p_j\rangle = 1-cos\_sim\langle x_{target},p_j\rangle,
\end{aligned}
\end{equation}
where cos\_sim is defined as above. $x_{target}$ is the sample to be identified and $p_j$ is the retrieved proxy. Then the label scores $score\_proxy$ are calculated as:
\begin{equation}
\begin{aligned} 
score\_proxy_j = \begin{cases}
\frac{1}{cos\_dis\langle x_{target},p_j\rangle} & p_j \in R_{proxy}\\
0 & p_j \notin R_{proxy}
\end{cases},
\end{aligned}
\end{equation}
where $R_{proxy}$ is the set of retrieved proxies.
Proxies closer to the target sample contribute higher scores, while proxies outside the retrieved set contribute a score of 0.
% Note that, proxies are single-labeled and each retrieved proxy contributes a certain score to its corresponding label.

Similarly, the sample-based k-NN retrieves the nearest $b$ samples and calculates the sample-sample distance:
\begin{equation}
\begin{aligned}
cos\_dis\langle x_{target},x_i\rangle = 1-cos\_ sim\langle x_{target},x_i\rangle,
\end{aligned}
\end{equation}
where $x_i$ is the retrieved sample.
The samples are associated with multiple labels, and each retrieved sample contributes the same score for the corresponding labels.
Then the label scores $score\_sample$ can be calculated by summing the contributions from all retrieved samples:
\begin{equation}
\begin{aligned}
score\_sample_j = \sum_{x_i\in R_{sample} \land y_{ij}=1} \frac{1}{cos\_dis\langle x_{target},x_i\rangle},
\end{aligned}
\end{equation}
where $R_{sample}$ is the set of retrieved samples.

At last, we combine the results of these two classifiers, \ie for label j, the total score is the weighted sum of these two terms:
\begin{equation}
\begin{aligned}
score_j = score\_proxy_j + \theta \times score\_sample_j,
\end{aligned}
\end{equation}
where $\theta$ is a hyperparameter that adjusts the weights of these two scores.
By combining the results of these two classifiers, our webpage identification considers both the uniform characteristics of different webpages and the variability of multi-tab traffic samples.
The scores for different webpages are aggregated and ranked, with the identified webpages being output based on a preset threshold $\tau$.

%-------------------------------------------------------------------------------
\section{Evaluation}

In this section, we evaluate \ours with datasets collected in the real world. 
We compare the performance of \ours with the state-of-the-art WF attacks.

\subsection{Experimental Setup}
\noindent \textbf{Implementation.}
We prototype \ours using Torch 1.9.0 and Python 3.8.
\zxy{We perform a random search of hyperparameters and set the optimal hyperparameters to default values as shown in Table~\ref{tab:parameter}.}
For the data augmentation module, we set the exchanging ratio $m_e$ to 5\% to sufficiently augment the original traffic without corrupting the critical traffic patterns.
For the feature transformation module, we use the weight of the sample-based loss to a larger value (\ie $\beta=4.5$) to enhance the quality of the transformed feature space under the multi-tab setting.
For the webpage identification module, we set the value of the threshold $\tau$ to achieve the best F1-score.
Further analysis of the impact of hyperparameters can be found in Section~\ref{sec:parameter_analysis}.

\begin{table}[t]
% \small
\centering
\caption{Hyperparameter settings %configurations 
in our evaluation.}
\small
\label{tab:parameter}
%\resizebox{0.49\textwidth}{!}{%
\begin{tabular}{c|c|c}
\toprule
\textbf{Module} & \textbf{Hyperparameters} & \textbf{Value} \\ \midrule
\multirow{2}{*}{Data Augmentation} 
& Input Dimension $d_i$ & 10,000 \\
& Exchanging Ratio $m_e$ & 5\% \\
\midrule
\multirow{3}{*}{Feature Transformation}
& Margin & 0.1 \\
& Loss Weight $\beta$ & 4.5 \\
& Transformed Dimension $d_o$ & 512 \\
\midrule
\multirow{3}{*}{Webpage Identification} 
& Neighbor Number $b$ & 40 \\
& Score Weight $\theta$ & 2 \\
& Threshold $\tau$ & 0.3 \\
\bottomrule
\end{tabular}
%}
\end{table}

\noindent\textbf{Dataset.}
Existing datasets~\cite{DF, AWF, ARES} regard each website as a distinct class.
To evaluate the performance of \ours on realistic WPF attacks, we collect multi-tab webpage traffic datasets.
To the best of our knowledge, these are the first datasets of real-world traffic from multi-tab webpages, where each webpage is regarded as a distinct class.
To be specific, we collect two multi-tab datasets: closed-world dataset and unmonitored webpage dataset under the open-world setting, as shown in Table~\ref{tab:dataset}.
\begin{itemize}[leftmargin=*]
    \item \textbf{Closed-World Dataset $\boldsymbol{CW}$:} We first build our monitored webpage set. Specifically, we select 115 websites from Alexa-top 20,000, visit the homepage of these websites and obtain 10 subpages by crawling the links on each website. Then we record the screenshots of these webpages during data collection, and filter out the invalid webpages by checking whether the screenshots exist and whether the webpage contents are successfully loaded. In this way, we acquire 1,000 webpages in total, and regard them as monitored webpages. We then collect samples of browsing the above 1,000 monitored webpages. The number of webpages visited in a session ranges from 1 to 5, with intervals between webpages randomly set between 3 and 10 seconds. 10 samples are collected for each webpage combination. We filter out samples with less than 1,000 packets and consider them as invalid accesses.
    %In total, we obtain 81,284 samples and the sample numbers with 1, 2, 3, 4, and 5 labels are 13,319, 28,210, 18,255, 12,491, 9,009.
    \item \textbf{Unmonitored Webpage Dataset under the Open-World Setting $\boldsymbol{OW}$:} For the unmonitored webpage set, we remove the websites that have been used by the closed world in Alexa-top 20,000 and keep the homepage of the rest websites.
    After filtering out invalid webpages, we get 9,236 webpages \zxy{from distinctive websites} and regard them as unmonitored webpages. We then collect samples of browsing a mixture of monitored webpages and unmonitored webpages. The number of webpages visited in a session is in the range of 2-5, with one from unmonitored webpages and the rest from monitored webpages. Each combination within this dataset consists of different unmonitored webpages and one sample is collected for each combination.
    \zxy{In this way, we ensure that the open-world traffic in the training set, validation set, and testing set originates from different unmonitored websites.}
    %In total, we obtain 9,236 samples of browsing both monitored and unmotored webpages under the open-world setting.
\end{itemize}

\begin{table}[t]
\small
    \centering
    \caption{Details of our datasets.}
    %\vspace{-0.1cm}
    \label{tab:dataset}
    \begin{threeparttable}
    \begin{tabular}{c|c|c|c|c}
    \toprule
       \textbf{Dataset} & \textbf{\makecell{Webpage \\ Number}} & \textbf{\makecell{Label / \\ Sample\tnote{1}}} &  \textbf{\makecell{Sample / \\ Comb.\tnote{2}}} & \textbf{\makecell{Sample \\ Number}} \\
    \midrule
        $\boldsymbol{CW}$ & 1,000 & 1-5 & 10 & 81,284 \\
        $\boldsymbol{OW}$ & 9,236 & 2-5 & 1 & 9,236 \\
    \bottomrule
    \end{tabular}
    \begin{tablenotes}    
        \footnotesize              
    \item[1] \textbf{Label / Sample} represents the number of labels per sample.
        \item[2] \textbf{Sample / Comb.} represents the number of samples per webpage combination.
  \end{tablenotes}
   \end{threeparttable}
\end{table}

The differences between our datasets and previous datasets are as follows: \first We separate the traffic of different subpages from the same website and treat each webpage as a distinct class, 
which is different from existing works that merely collect traffic of index pages~\cite{AWF, DF} or utilize subpage traffic to identify websites~\cite{ARES, gandalf, sok, goodenough}.
\second We expand the scale of monitored webpages. Most previous works only collect and monitor around 100 websites ~\cite{DF, TF, ARES, MWF, BAPM},
\zxy{whereas we monitor 1,000 webpages in total.}
\zxy{Specifically, our datasets are the largest multi-tab webpage datasets in the wild.}
\third In the real world, the number of webpages visited by clients in a session is dynamic, which is not fully considered by existing works~\cite{BAPM}. To adapt to this dynamism, the label number of samples in our datasets is not fixed.

\noindent\textbf{Baselines.} We compare \ours with six state-of-the-art WF attacks, divided into two categories.
{
\begin{itemize}[leftmargin=*]
\item \textbf{Single-Tab Attacks.} We select one machine-learning-based method k-FP~\cite{kFP} and three deep-learning-based methods DF~\cite{DF}, Tik-Tok~\cite{Tik-Tok} and NetCLR~\cite{2023aug}. k-FP extracts 175 features from time and direction sequences and applies the Random Forest classifier for website classification.
The original k-FP attack uses 1,000 trees, which cannot be realized due to the high resource overhead under the large-scale multi-tab setting. Therefore, we have to reduce the tree number to 100. 
DF and Tik-Tok adopt CNN-based architecture to achieve automatic feature extraction.
NetCLR applies self-supervised learning to pre-train a DF model based on augmented traces and finetunes the model with labeled traces.
Since the original cross-entropy loss adopted by the above three methods is suitable for multi-class classification, we follow~\cite{ARES} and change their loss function to the binary cross-entropy loss to achieve multi-label classification. 
\item \textbf{Multi-Tab Attacks.}
We select two multi-tab attacks BAPM~\cite{BAPM} and TMWF~\cite{tmwf}.
BAPM applies multi-head attention for multi-tab identification and each head predicts an individual website.
TMWF applies the DETR algorithm in object detection and each query predicts a website.
Since BAPM can only identify a fixed number of webpages, we follow~\cite{tmwf} and set a "no-tab" class for these two methods. We set N attention heads for BAPM and N tab queries for TMWF where $N=5$ since the maximum number of concurrently accessed webpages in our datasets is 5.
\end{itemize}
}

Due to various practical limitations, we do not compare \ours with all previous attacks.
Specifically, TF~\cite{TF} mines samples based on single-label samples, leading to the problem of class collapse under the multi-label setting.
MWF~\cite{MWF, MWF2} can only identify the first webpage under the multi-tab setting.
ARES~\cite{ARES} builds a separate, complex Transformer-based classifier for each website, making it impractical for large-scale webpage fingerprinting due to its high overhead.
Therefore, these attacks are excluded from our experiments.

\noindent\textbf{Evaluation Metrics.} We use two multi-label classification metrics for evaluation: Recall@k and AP@k~\cite{metric, metric2}.
Rather than assess the result with the highest predicted probability, Recall@k evaluates the recall rate of the top k predicted webpages with the highest probabilities.
Specifically, for sample $i$, assuming the real set of visited webpages is $y_i$, and the set of top-$k$ predicted webpages is $\hat{y_i}$, Recall@k is calculated as follows:
\begin{equation}
\begin{aligned}
Recall@k =\frac{\|y_i \cap \hat{y_i}\|}{\|y_i\|}.
\end{aligned}
\end{equation}

AP@k is the average of Precision@k, which measures the proportion of correctly predicted webpages among the top-$k$ results.
Since the number of visited webpages in our datasets varies, AP@k can reflect the performance more precisely. Specifically, for sample $i$, assuming the real set of visited webpages is $y_i$, AP@k can be calculated as:
\begin{equation}
\begin{aligned}
AP@k  =\frac{\sum_{t=1}^{k} Precision@t}{min(k,\|y_i\|)}.
\end{aligned}
\end{equation}
To compute Precision@t, we get the top-$t$ predicted webpages with the highest probabilities $\hat{y_i}$. Precision@t is then defined as:
\begin{equation}
\begin{aligned}
Precision@t =\frac{\|y_i \cap \hat{y_i}\|}{t}.
\end{aligned}
\end{equation}

Note that we do not calculate averaged Recall@k because the denominator of Recall@k is the number of the ground truth label of the sample, which is the same under different $k$ values. We calculate Recall@k and AP@k for each sample in the dataset, and report the average as the final results.

\subsection{WPF Attacks in the Closed World}
We first evaluate the performance of \ours under the closed-world setting, where clients only visit monitored webpages and the attacker can collect traffic samples of all the webpages to train the model.
We use our closed-world dataset $\boldsymbol{CW}$ for evaluation.
We divide the dataset into the training, validation and testing sets with the ratio of 8:1:1.
\zxy{Specifically, we utilize the augmented training set to train the feature transformation model and finetune parameters on the validation set. Then we use the trained model to transform samples in the testing set and achieve webpage identification based on the updated proxies and transformed training samples.}
For NetCLR, we use the training set to pre-train the model and the validation set to finetune.
We calculate the multi-label classification metrics Recall@k and AP@k with the $k$ values of Recall@k in $\{5,10, 15, 20, 25, 30\}$, and AP@k in $\{1, 2, 3, 4, 5\}$.
Since BAPM and TMWF predict webpages based on each attention head or tab query and are unable to determine the probability of all the webpages visited in a session, we can only compare with them on Recall@5.

\begin{figure}
        \begin{minipage}[t]{1\linewidth}
		\centering
		\includegraphics[width=2.6in]{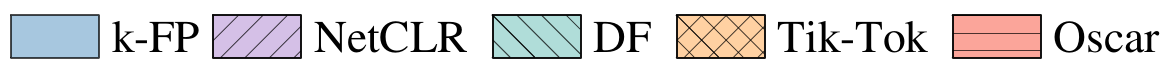}
	\end{minipage}
	\begin{minipage}[t]{0.48\linewidth}
		\centering
		\includegraphics[width=1.6in]{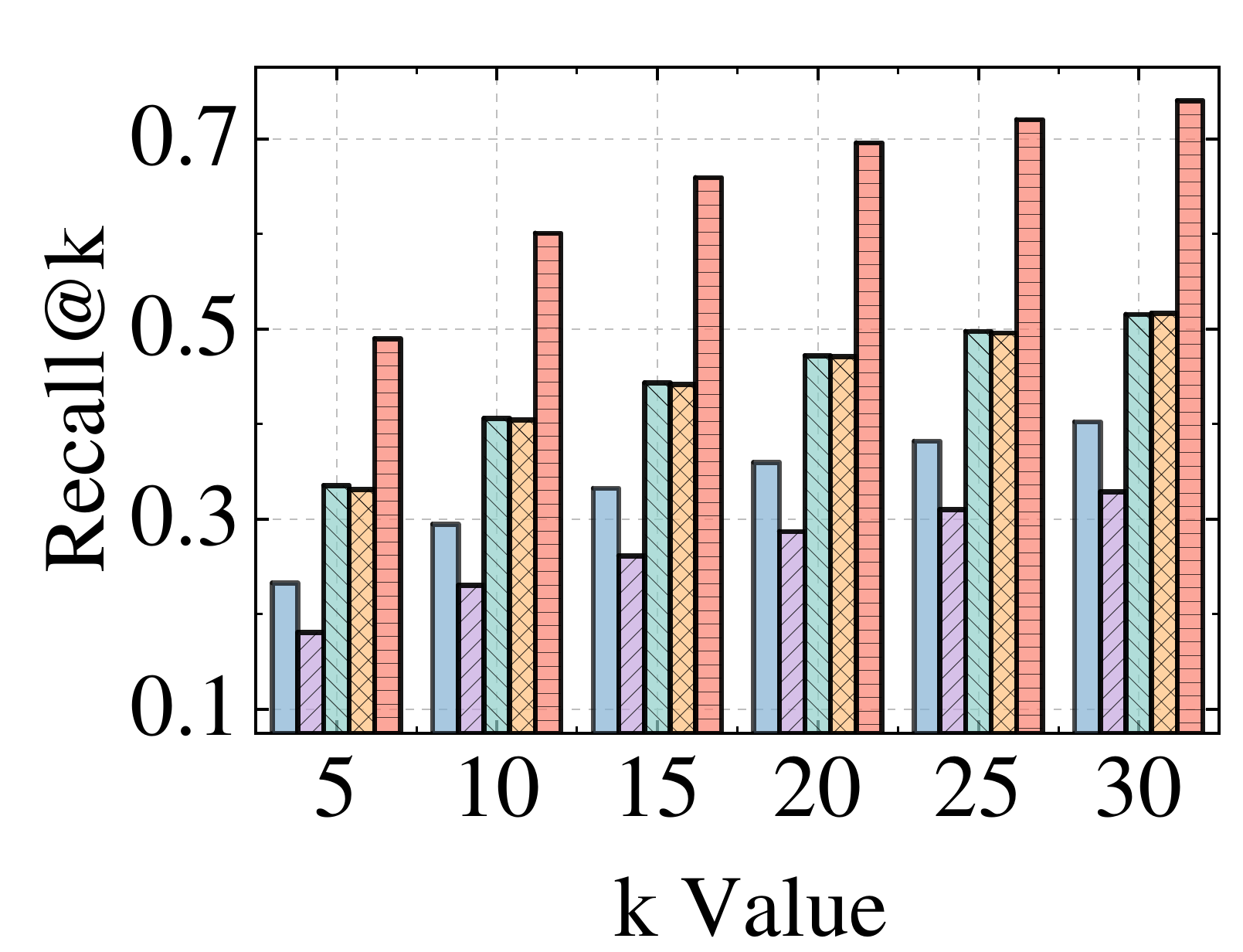}
	\end{minipage}
	\begin{minipage}[t]{0.48\linewidth}
		\centering
		\includegraphics[width=1.6in]{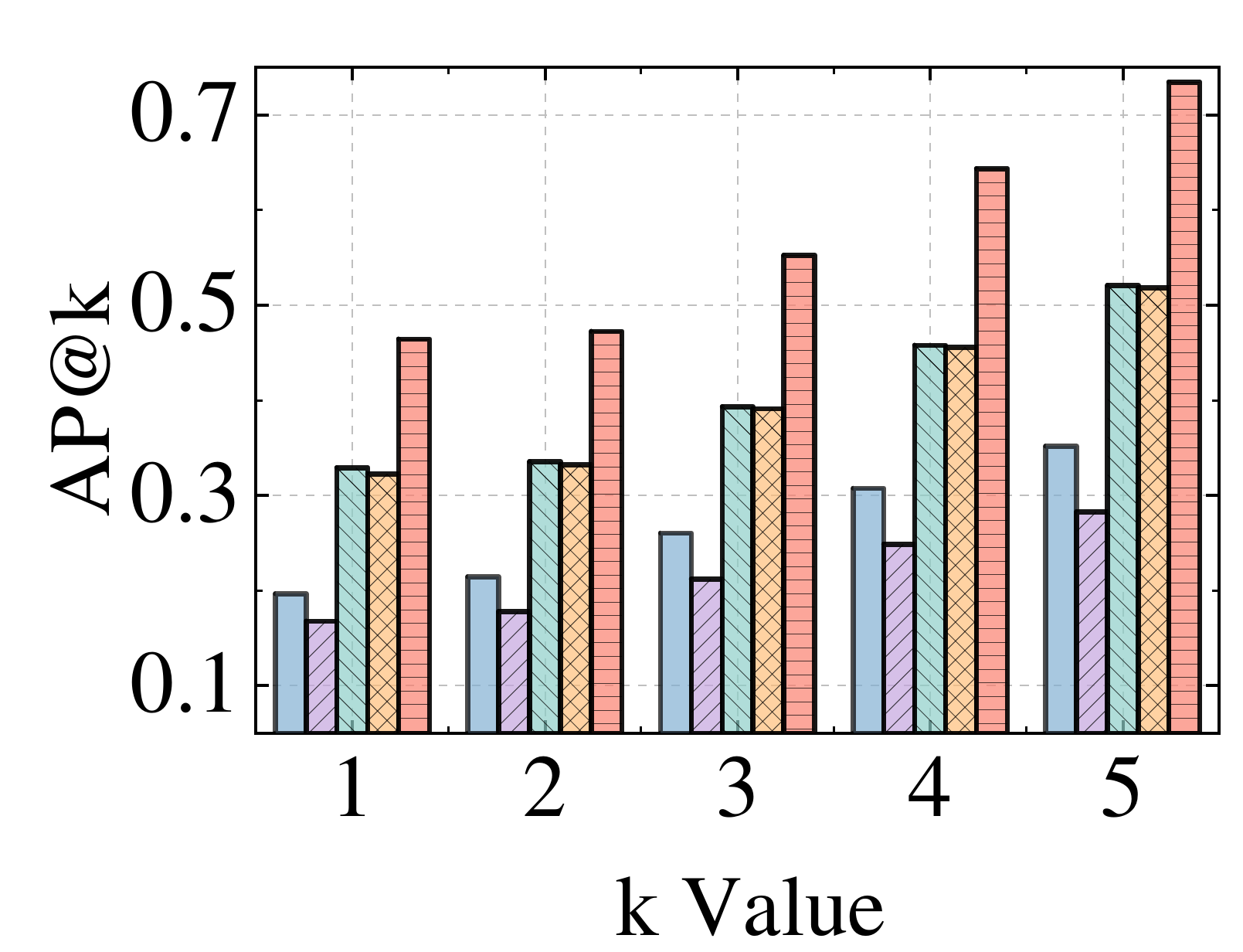}
	\end{minipage}
	\caption{Results of the closed-world experiment. We report the Recall@k and AP@k with different $k$ values.}
	\label{fig:closed-world}
\end{figure}

We present the Recall@k and AP@k of five attacks in Figure~\ref{fig:closed-world} and Recall@5 of all methods in Table~\ref{tab:closed-world}.
Results show that Recall@k and AP@k increase as the k value increases and \ours achieves the best performance in all metrics. %in all Recall@5 to Recall@30 and AP@1 to AP@5.
Specifically, Recall@30 and AP@5 of \ours are both over 0.73, while the best results of other methods remain around 0.52.
% \zxy{Compared with SOTA methods, \ours reaches an average Recall@5 improvement of }
Compared with k-FP, NetCLR, DF, Tik-Tok, BAPM and TMWF, \ours improves by 110.2\%, 170.8\%, 46.1\%, 47.9\%, 132.6\% and 24.0\% on Recall@5.
The performance superiority demonstrates that \ours can identify the webpages both comprehensively and accurately. %, but its results are also more credible.
Although existing attacks achieve good performance in terms of website identification, their effectiveness significantly declines when applied to multi-tab webpage identification.
This performance reduction is primarily due to their incapability of capturing the subtle distinctions hidden within the high-dimensional feature vectors of similar webpages.
In contrast, \ours concentrates on analyzing the differences among various webpages through webpage comparison.
The proxy-based metric learning loss effectively clusters traffic from the same webpages and the sample-based metric learning loss isolates traffic from irrelevant webpages.
In this way, we separate different webpages and extract their distinct characteristics in the transformed feature space.
In addition, the improved multi-label webpage identification comprehensively takes into account the uniform characteristics of webpages and the diversity of multi-tab traffic, therefore achieving more accurate results under the multi-tab setting.

\begin{table}[t]
\small
\setlength\tabcolsep{2.5pt}
    \centering
    \caption{Recall@5 of different attacks under the closed-world setting.}
    %\vspace{-0.1cm}
    \label{tab:closed-world}
    \begin{tabular}{c|c|c|c|c|c|c|c}
    \toprule
        \textbf{Attacks} & k-FP & NetCLR & DF  & Tik-Tok & BAPM & TMWF & \ours \\
    \midrule
       \textbf{Recall@5} & 0.2331 & 0.1809 & 0.3354 & 0.3313 & 0.2106 & 0.3951 & \textbf{0.4899} \\
    \bottomrule
    \end{tabular}
\end{table}

\noindent \textbf{Remark.} In a nutshell, \ours achieves the best performance in identifying multiple webpages from obfuscated traffic, which is credited to the effectively transformed features based on our method.
Besides, the proxy-based and sample-based combined webpage identification can classify webpages more accurately.

\subsection{WPF Attacks in the Open World}
Under the open-world setting, the attacker can only collect samples from a subset of webpages, which does not cover all the webpages in the testing set. 
{Typically, sensitive webpages constitute only a fraction of the entire webpages, and our primary objective is to precisely identify the particular monitored sensitive webpages despite the interference of unmonitored webpages.}
Under the open-world setting, following existing works\cite{ARES, DF}, we mix the closed-world dataset $\boldsymbol{CW}$ and the unmonitored webpage dataset in the open world $\boldsymbol{OW}$ to ensure that the sample number of monitored and unmonitored webpages is balanced.
All the unmonitored webpages are treated as a single class, while each monitored webpage is still regarded as a distinct class.
Note that samples under the open-world setting are still multi-labeled and the number of sample labels is also dynamic.

\begin{figure}
        \begin{minipage}[t]{1\linewidth}
		\centering
		\includegraphics[width=2.6in]{figures/cutline1.pdf}
	\end{minipage}
	\begin{minipage}[t]{0.48\linewidth}
		\centering
		\includegraphics[width=1.6in]{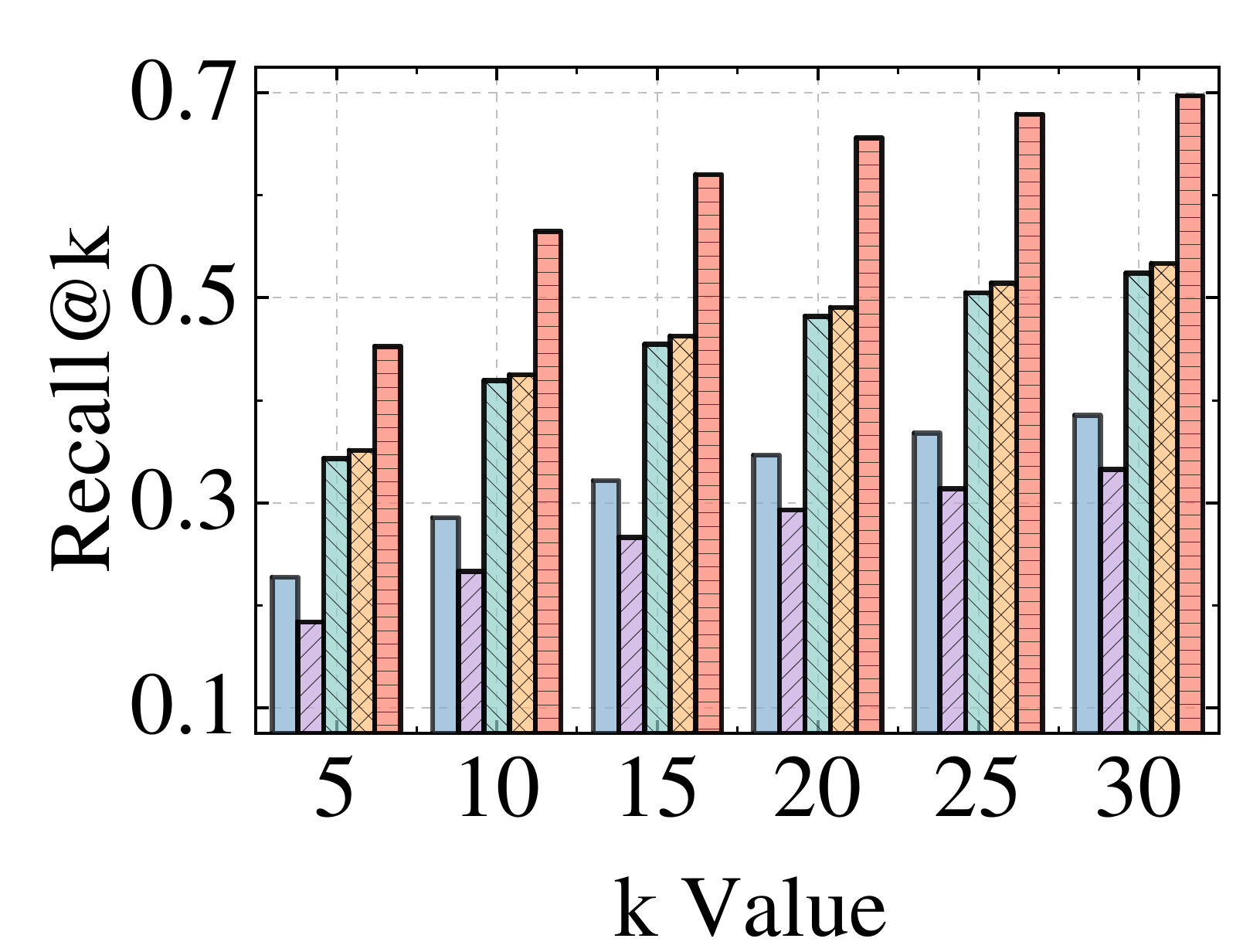}
	\end{minipage}
	\begin{minipage}[t]{0.48\linewidth}
		\centering
		\includegraphics[width=1.6in]{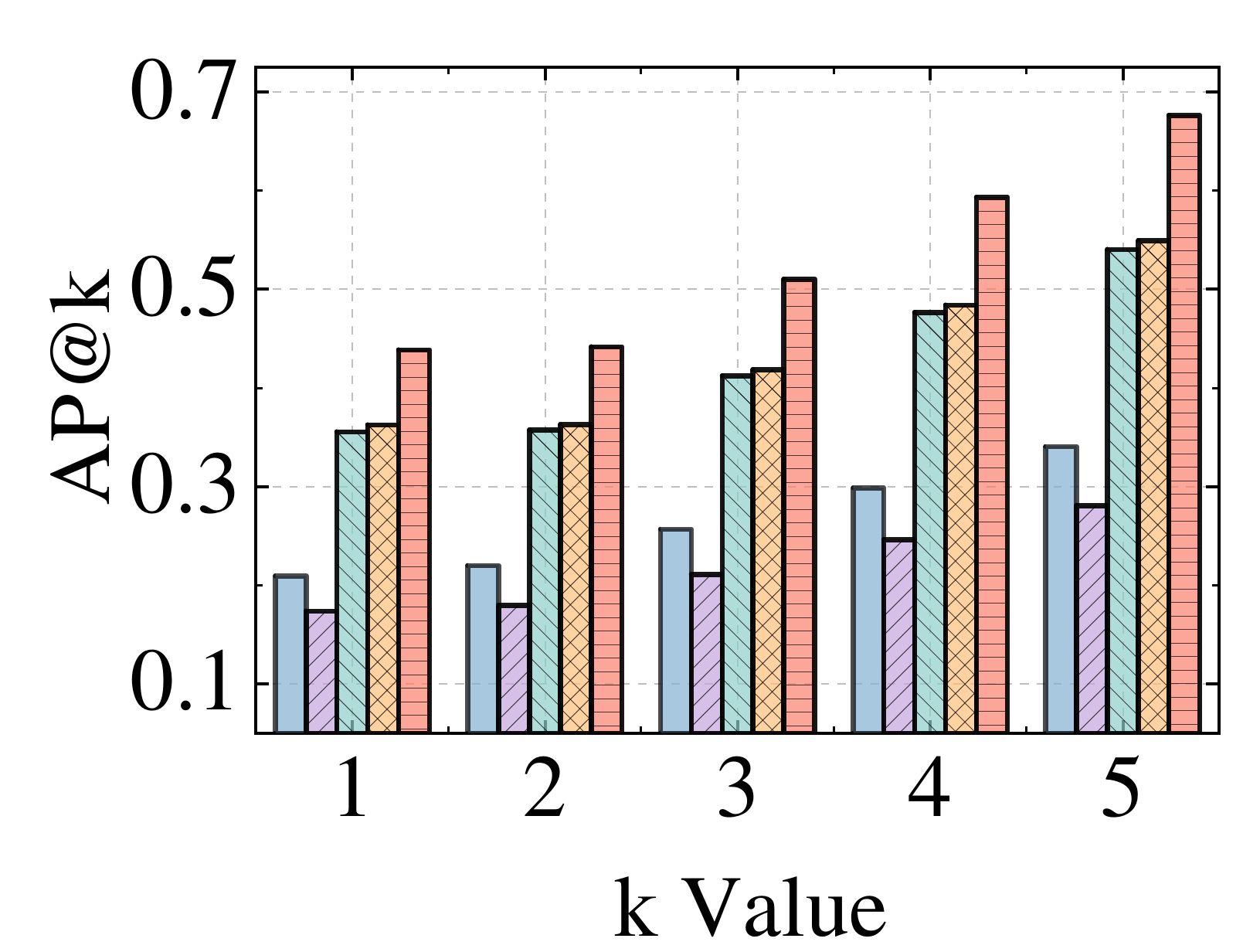}
	\end{minipage}
	\caption{Results of the open-world experiment. We report the Recall@k and AP@k with different $k$ values.}
	\label{fig:open-world}
\end{figure}

Figure~\ref{fig:open-world} shows the performance under the open-world setting.
{Since BAPM and TMWF cannot distinguish between the unmonitored and the padding webpages under the open-world setting, we do not compare with them.}
Results show that Recall@30 of \ours remains around 0.7 and AP@5 remains over 0.67.
Specifically, \ours improves by an average of 63.5\% and 72.0\% on Recall@30 and AP@5 respectively.
% Specifically, \ours reaches a Recall@5 improvement of 99.2\%, 146.6\%, 31.9\% and 28.9\%, and AP@5 improvement of 98.6\%, 141.1\%, 25.2\%, 23.2\% over k-FP, NetCLR, DF, and Tik-Tok, respectively.
Therefore, \ours can still identify webpages more accurately than existing attacks in the presence of a large number of unmonitored webpages.
This is mainly due to our feature transformation design. The proxy-based loss can individually cluster traffic from monitored webpages and unmonitored webpages, and the sample-based loss can further isolate irrelevant traffic.

Note that there is a slight performance drop compared to the closed-world setting.
We believe this is mainly because traffic patterns of different unmonitored webpages exhibit obvious variations.
When they are regarded as the same class in the training phase, it is challenging to extract the common feature pattern of all these webpages.
In addition, the traffic of unmonitored webpages in the testing phase does not necessarily share similar patterns with the unmonitored webpages in the training phase, and may even resemble monitored webpages more closely, which brings confusion to the model.
In spite of this, \ours still outperforms previous works under the open-world setting by separating traffic from various monitored webpages and unmonitored webpages.

\noindent \textbf{Remark.} Overall, \ours achieves the best performance under the open-world setting, where traffic of monitored webpages is mixed with that of unmonitored webpages. This demonstrates the superiority of \ours in handling real-world attacks with a substantial number of webpages.

\subsection{WPF Attacks on Various Scales of Webpages}
We further evaluate the performance of \ours on various scales of monitored webpages. We consider the large-scale evaluation and set the size of webpages as 700, 800, 900, 1,000.
We use the closed-world dataset $\boldsymbol{CW}$ for this experiment and the details of datasets with different webpage numbers are shown in Table~\ref{tab:small-scale data}.
The datasets are constructed as follows: we first randomly sample different numbers of webpages from the full set of monitored webpages, and then filter traffic samples from the original dataset.
We keep the samples whose labels overlap with the selected webpages, and leave out the others.
For those selected samples, the labels corresponding to the selected webpages are retained, and other labels are ignored. 
We use Recall@5 and AP@5 as metrics to evaluate the performance of different attacks.

\begin{table}[t]
\small
    \centering
    \caption{{Details of the datasets for the evaluation on different scales of monitored webpages.}}
    %\vspace{-0.1cm}
    \label{tab:small-scale data}
    \begin{tabular}{c|c}
    \toprule
       \textbf{Webpage Number} & \textbf{Sample Number} \\
    \midrule
        700 & 70,889 \\ 
        800 & 74,683 \\
        900 & 77,387 \\
        1,000 & 81,284 \\
    \bottomrule
    \end{tabular}
\end{table}

\begin{figure}
        \begin{minipage}[t]{1\linewidth}
		\centering
		\includegraphics[width=2.2in]{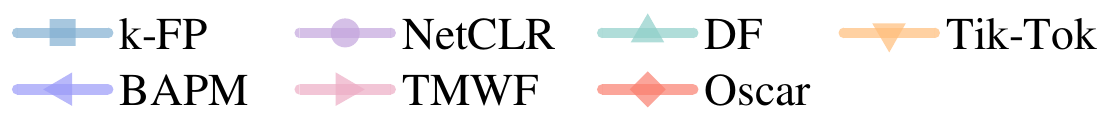}
	\end{minipage}
	\begin{minipage}[t]{0.48\linewidth}
		\centering
		\includegraphics[width=1.58in]{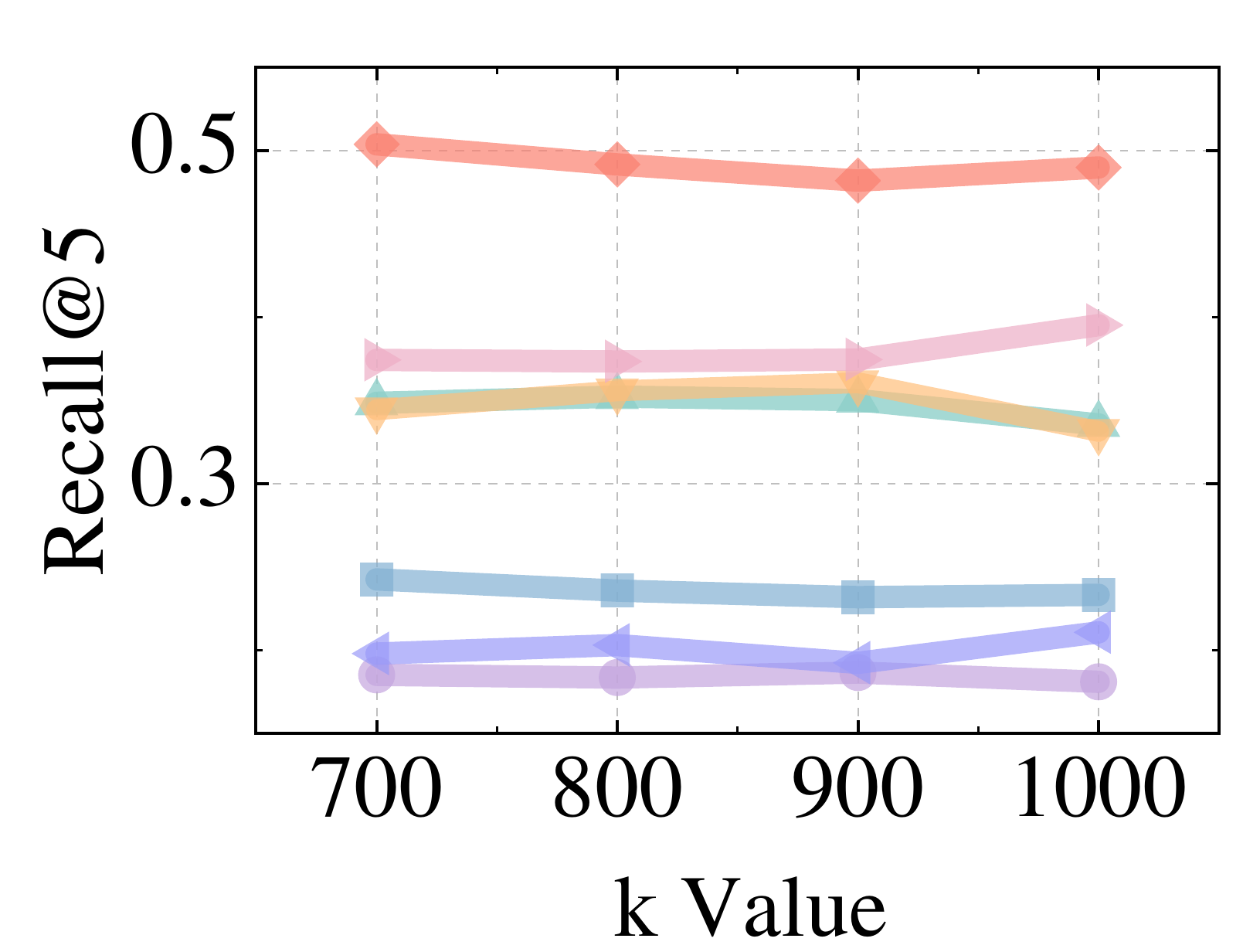}
	\end{minipage}
	\begin{minipage}[t]{0.48\linewidth}
		\centering
		\includegraphics[width=1.6in]{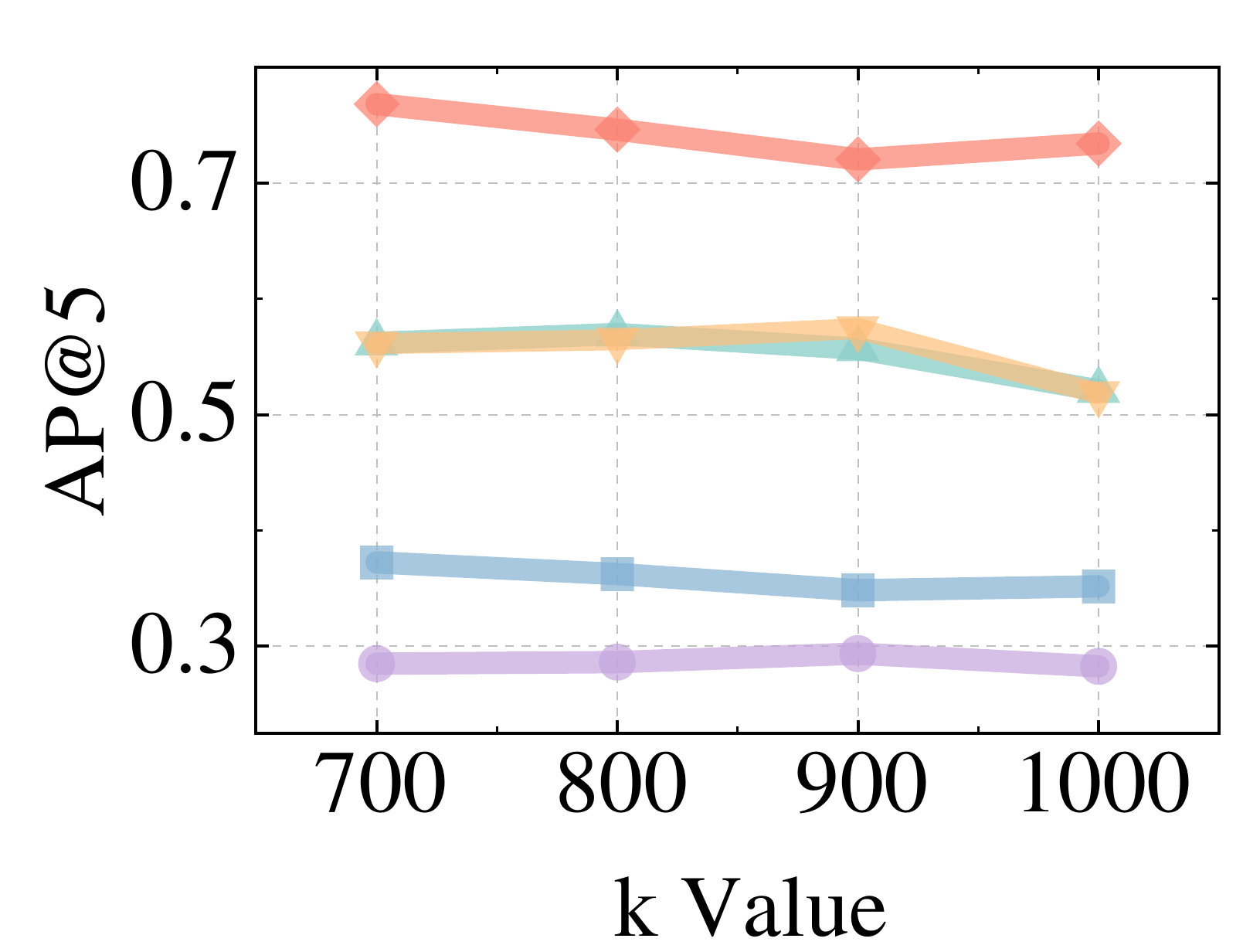}
	\end{minipage}
	\caption{Results of the experiment on various scales of webpages, where the webpage number ranges from 700 to 1,000.}
	\label{fig:small-scale}
\end{figure}

The results are presented in Figure~\ref{fig:small-scale}.
As mentioned above, BAPM and TMWF cannot calculate AP@5, so we do not compare with them on this metric.
Results demonstrate that \ours consistently delivers superior performance across various scales of webpages.
Specifically, AP@5 of \ours is maintained over 0.72, while the best result of existing attacks is less than 0.58.
\zxy{Furthermore, Recall@5 of \ours declines by only 2.76\% and AP@5 declines by 4.41\% as the webpage number increases from 700 to 1,000.}
This demonstrates the potential of \ours to maintain effective across different scales of monitored webpages.
As the number of webpages grows, the challenge to distinguish them in the original feature space increases.
But \ours focuses on contrasting webpages against each other to extract the subtle differences among them.
As a result, \ours can separate different webpages even if they are hard to distinguish in the original feature space.
Furthermore, the multi-label classification based on k-NN is efficient and effective, allowing realistic large-scale attacks in the wild.

\noindent  \textbf{Remark.} \ours outperforms existing attacks in classifying various scales of webpages with the performance fluctuation kept within a manageable range. 
When the scale of monitored webpages further rises, \ours is expected to sustain efficient and accurate attacks due to its ability to extract the differences among webpages.

\subsection{Ablation Study}
Next, we conduct the ablation study to analyze the impact of each module and the two loss functions on the attack performance. The experiment is conducted under both the closed- and open-world settings and we report Recall@5 and AP@5. We consider the following five settings: \first k-NN classifier with the raw features; \second k-NN classifier with proxy-based and sample-based combined feature transformation; \third k-NN classifier with data augmentation and proxy-based feature transformation; \fourth k-NN classifier with data augmentation and sample-based feature transformation; \fifth k-NN classifier with data augmentation and proxy-based and sample-based combined feature transformation.

\begin{table}
\small
  \centering
  \caption{Ablation study results.}
  \label{tab:ablation}
  \begin{threeparttable}
  \scalebox{0.87}{
  \begin{tabular}{c|c|c|c|c}
   \toprule
    \textbf{\multirow{2}{*}{Method}}  & \multicolumn{2}{c|}{\textbf{Closed-World}}  & \multicolumn{2}{c}{\textbf{Open-World}} \\
    \cmidrule{2-5}
    
    & \textbf{Recall@5} & \textbf{AP@5}  & \textbf{Recall@5} & \textbf{AP@5} \\
    \midrule
    WI\tnote{1} & 0.0155 & 0.0189 & 0.0238 & 0.0234 \\
    FT\tnote{3} \ (combined)\ +WI\tnote{1} & 0.4511 & 0.6749 & 0.4206 & 0.6272 \\
    DA\tnote{2}\ +FT\tnote{3} \ (proxy-based)\ +WI\tnote{1} & 0.3066 & 0.4450 & 0.2996 & 0.4340 \\
    DA\tnote{2}\ +FT\tnote{3} \ (sample-based)\ +WI\tnote{1} & 0.0063 & 0.0070 & 0.0413 & 0.0826 \\
    DA\tnote{2}\ +FT\tnote{3} \ (combined)\ +WI\tnote{1} & \textbf{0.4899} & \textbf{0.7344} & \textbf{0.4527} & \textbf{0.6766} \\
    \bottomrule
  \end{tabular}}
  
  \begin{tablenotes}    
        \footnotesize              
        \item[1] \textbf{WI} represents the webpage identification module. 
        \item[2] \textbf{DA} represents the data augmentation module. 
        \item[3] \textbf{FT} represents the feature transformation module. 
  \end{tablenotes}
  \end{threeparttable}
\end{table}

The experiment results are detailed in Table~\ref{tab:ablation}.
When k-NN classifies based on the original feature space, it achieves a poor performance in distinguishing different webpages.
This is due to the stochastic distributions of webpage traffic in the original feature space.
The incorporation of the feature transformation module greatly improves the identification performance of \ours, as it transforms the feature space to cluster samples of the same webpages and separate samples of different webpages.
Additionally, the data augmentation module generates more samples and enhances the sample diversity, bringing further performance improvement.

Regarding the metric learning loss functions, the vanilla proxy-based loss underperforms the combined loss as it ignores the relationship among webpages under the multi-label setting.
Besides, the vanilla sample-based loss aims at isolating irrelevant webpage traffic but cannot cluster relevant webpage traffic, making it ineffective when used alone.
Nevertheless, it can significantly improve the performance when combined with the proxy-based loss, as it contributes to the optimization of webpage distributions in the feature space by isolating irrelevant webpage traffic.

\noindent \textbf{Remark.} To summarize, the feature transformation module plays a vital role in webpage identification by effectively separating traffic of different webpages in the transformed feature space. The data augmentation module boosts the performance by generating more samples and enhancing the sample diversity. The proxy-based loss contributes to aggregate traffic from the same webpages, and the sample-based loss is pivotal in separating irrelevant traffic. Therefore, combining these two losses results in the best performance.

%-------------------------------------------------------------------------------
\subsection{Analysis of Hyperparameters}
\label{sec:parameter_analysis}
%-------------------------------------------------------------------------------
\begin{figure}[t]
  \centering
  \begin{minipage}[t]{1\linewidth}
		\centering
		\includegraphics[width=1.4in]{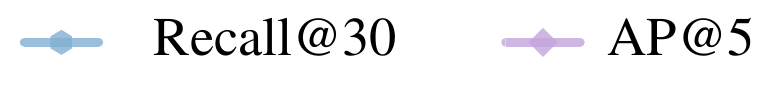}
	\end{minipage}
  \subfigure[Loss weight]{
    \label{fig:beta}
    \includegraphics[width=0.48\linewidth]{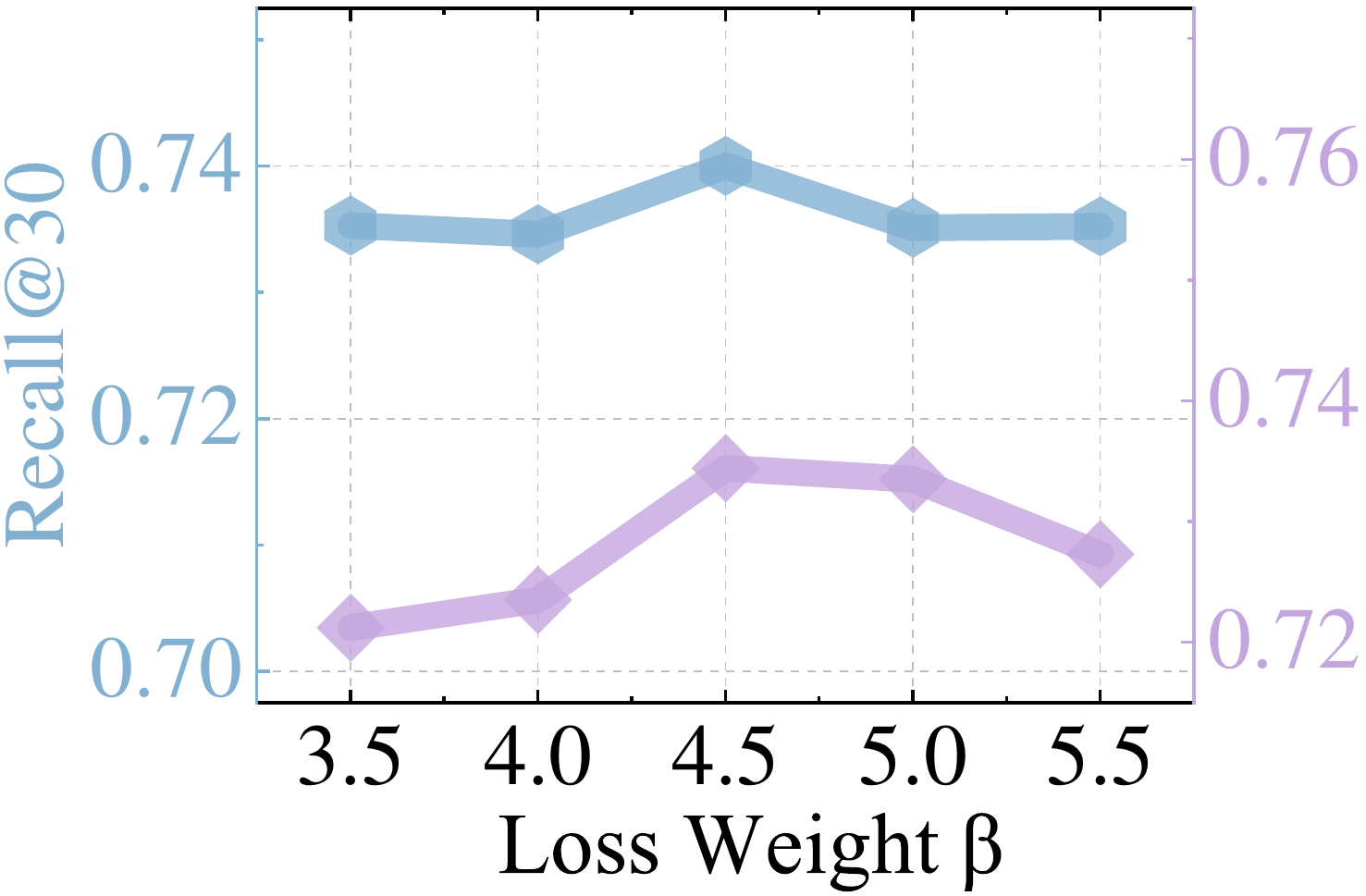}
  }
  \subfigure[Neighbor number]{
    \label{fig:neighbor_number}
    \includegraphics[width=0.46\linewidth]{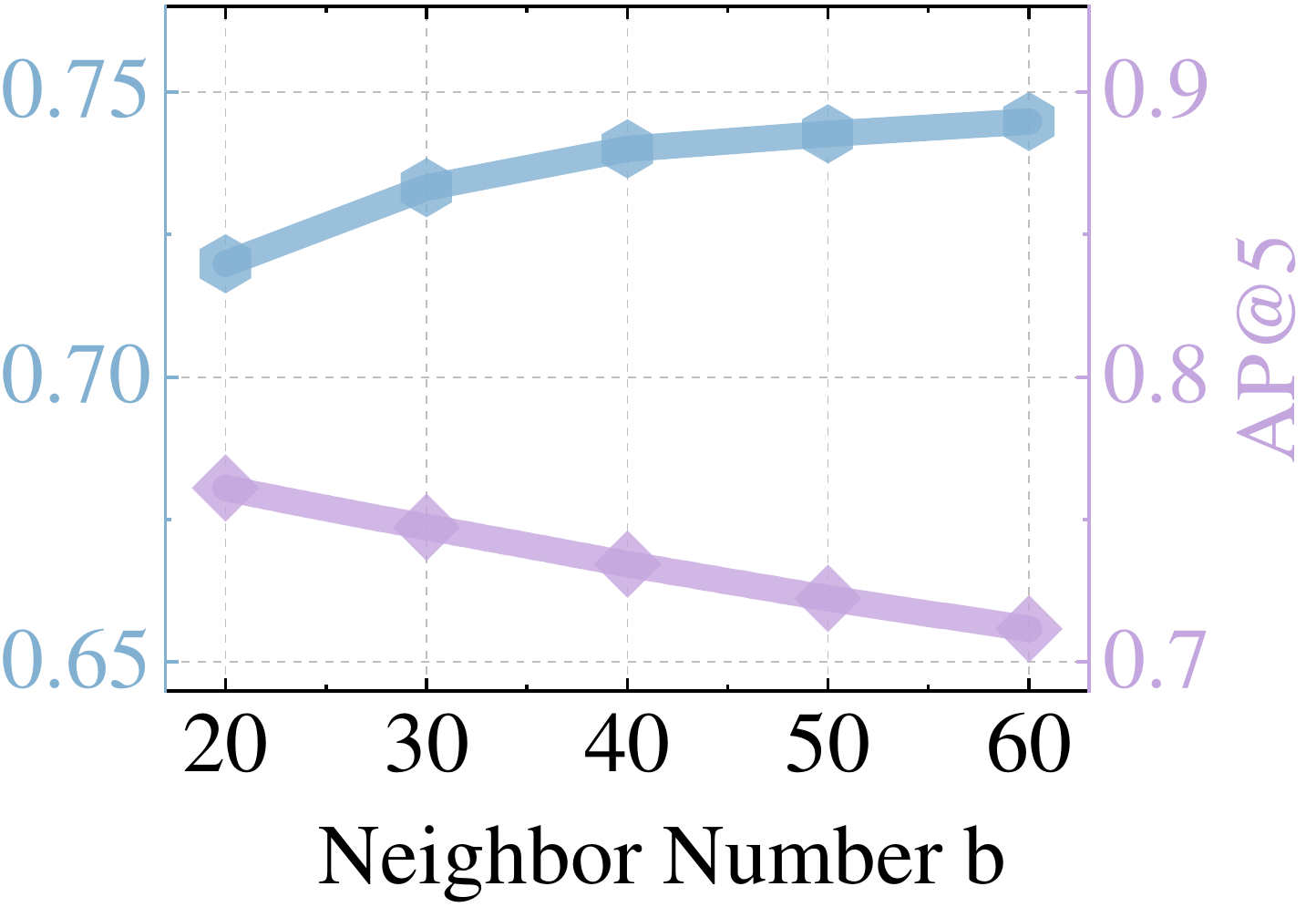}
  }
  \caption{Impacts of two critical hyperparameters on the performance: loss weight $\beta$ in the Feature Transformation module and neighbor number $b$ in the Webpage Identification module.}
  \label{fig:hyperparameter}
\end{figure}

In this section, we analyze the impacts of critical hyperparameters on the performance of \ours. We select two hyperparameters: loss weight $\beta$ in the Feature Transformation module and neighbor number $b$ in the Webpage Identification module. $\beta$ adjusts the weights of the proxy-based and sample-based losses, and $b$ decides the number of the retrieved neighbor proxies and samples. We evaluate the above two hyperparameters in the closed world, and all other hyperparameters are set to the default setting when evaluating each hyperparameter.

Figure~\ref{fig:beta} shows the performance with different $\beta$ values.
The fluctuations in Recall@30 and AP@5 are maintained within 0.015, demonstrating a stable performance with different settings.
We note that \ours achieves better performance when assigning a greater weight to the sample-based loss, demonstrating the importance of separating irrelevant traffic under the multi-tab setting.
Through this separation, the sample-based loss contributes to extracting the relationship among webpages, therefore optimizing the distributions of different webpages in the transformed feature space.

Figure~\ref{fig:neighbor_number} demonstrates the impact of $b$ values on the performance.
When retrieving more proxies and samples, Recall@30 exhibits a slight increase while AP@5 shows a slight decrease.
But both metrics stay within a narrow range from 0.71 to 0.77.
Considering both the results of Recall and AP, we set the number of neighbor proxies and samples as 40.

\noindent\textbf{Remark.} Overall, the variation in hyperparameter settings for \ours demonstrates only a modest impact on its performance, underscoring that its superior performance is due to the robust design instead of hyperparameter settings. This stability across different settings highlights the model's adaptability and reliability, making it well-suited for real-world applications.

%-------------------------------------------------------------------------------
\section{Discussion}
%-------------------------------------------------------------------------------

\noindent\textbf{Larger-Scale of Monitored Webpages.} We assess the performance of \ours using 1,000 monitored webpages in the experiments, which is still quite limited compared to the actual number of webpages.
The substantial size of webpages leads to a larger number of potential webpage combinations, which makes it impossible to achieve a comprehensive analysis that matches the number of real-world webpages.
However, existing experiments still demonstrate the superiority of \ours on various scales of webpages and the potential to maintain effective on a even larger scale of webpages.
This effectiveness is largely attributed to our feature transformation design.
Despite that the increase in the webpage number naturally reduces the disparities in the original feature space, \ours employs comparisons among different webpages, enabling it to effectively discern and capture these subtle distinctions.
Therefore, it is capable of separating webpages in the transformed feature space even when the scale is further up.

\noindent\textbf{Deploying Complex Webpage Fingerprinting Defenses.} Recently, an increasing number of works on webpage fingerprinting defenses have been proposed. Existing defenses can be divided into the following five categories: molding traffic into fixed patterns~\cite{dyer2012peek, cai2014cs, cai2014systematic, lu2018dynaflow}, adding dummy packets~\cite{wtf-pad, gong2020zero}, creating collisions among webpages~\cite{nithyanand2014glove, wang2014effective, wang2017walkie}, splitting traffic into multiple streams~\cite{henri2020protecting, de2020trafficsliver} and introducing adversarial noise~\cite{rahman2020mockingbird, nasr2021defeating, gong2022surakav}. 
However, many existing defenses incur high latency and bandwidth overhead, making them impractical for real-world deployment~\cite{sok}.
We will focus on improving the robustness of WPF attacks under complex defenses in future work.

\noindent\textbf{Robustness under Concept Drift.} Concept drift is incurred by the discrepancy between training data and testing data as the webpage properties change over time~\cite{ARES}. In reality, the contents of webpages are constantly changing, leading to variations in traffic patterns.
Therefore, the trained model may not be well adapted to identify traffic a few years later, leading to a decrease in identification accuracy.
However, \ours learns the differences among webpages through metric learning instead of mapping features to specific labels.
\zxy{Thus, the framework based on feature transformation remains effective in discerning the differences among webpages, even amidst content changes.
In addition, we can finetune the feature transformation model and update the proxies by collecting a modest amount of new samples to enhance the accuracy under concept drift.}

%-------------------------------------------------------------------------------
\section{Related Work}
%-------------------------------------------------------------------------------

\noindent\textbf{Single-Tab WF Attacks.}
WF attacks compromise the online privacy of Tor clients by extracting website fingerprints from Tor traffic. 
Early attacks~\cite{panchenko2011website, wang2014effective, CUMUL, kFP} extract website fingerprints based on expert knowledge and utilize ML models for website identification. 
Recently, DL has been widely applied to enhance the performance of WF attacks. 
AWF~\cite{AWF} utilizes DL models for automatic feature extraction and analysis. 
DF~\cite{DF} develops an improved CNN model capable of robust WF attacks against the WTF-PAD defense~\cite{wtf-pad}.
Tik-Tok~\cite{Tik-Tok} and RF~\cite{shen2023subverting} improve traffic feature representations, further enhancing the robustness of WF attacks. 
However, the excellent performance of DL-based WF attacks depends on a large amount of training data.
Var-CNN~\cite{Var-CNN}, TF~\cite{TF} and GANDaLF~\cite{gandalf} improve the model architecture and training method to achieve effective WF attacks with a small number of training samples. 
NetCLR~\cite{2023aug} augments traces and applies self-supervised and semi-supervised learning to enhance the robustness across different network conditions.
\zxy{
Holmes~\cite{deng2024ccs} implements an early-stage WF attack by analyzing the spatio-temporal distribution features of traffic.
Mitseva et al.~\cite{mitsevastop} analyze traffic from multiple subpages of the same website to enhance website identification.}
Different from existing WF attacks that target on website identification, \ours effectively achieves a fine-grained webpage fingerprinting attack.

\noindent \textbf{WPF Attacks.}
Fine-grained WPF attacks present significant challenges because multiple subpages of a website often share similar templates and layouts, resulting in more similar traffic patterns~\cite{shen2020fine, wang2020high}.
Existing WPF attacks mainly identify webpages by extracting packet-level features and flow-level features~\cite{shen2019webpage, shen2020fine, zhang2019deep, shen2021efficient,lu2021gap}. 
For example, BurNet~\cite{shen2021efficient} extracts features from unidirectional burst sequences based on CNN, and GAP-WF~\cite{lu2021gap} utilizes GNN to extract flow-level features.
However, existing WPF attacks assume that Tor clients only open a single tab to access webpages. 
\ours relaxes the assumption of existing WPF attacks, achieving more realistic multi-tab WPF attacks. 
Even with the interference of noise packets from other webpages, \ours still achieves robust WPF attacks.

\noindent \textbf{Multi-tab WF Attacks.}
The obfuscated traffic under the multi-tab setting imposes challenges of extracting the pure traffic patterns of each website~\cite{MWF, CWF}. Existing multi-tab WF attacks identify obfuscated traffic under the multi-tab setting by applying the attention mechanism~\cite{BAPM, ARES, tmwf}. For instance, BAPM~\cite{BAPM} and ARES~\cite{ARES} use the multi-head attention mechanism, and TMWF~\cite{tmwf} integrates powerful Transformer models for feature extraction. 
However, existing multi-tab attacks aim at differentiating websites rather than fine-grained webpages and are only applicable to a small scale of monitored websites. \ours achieves a large-scale WPF attack under the multi-tab setting.

%-------------------------------------------------------------------------------
\section{Conclusion}
%-------------------------------------------------------------------------------
In this work, we propose \ours, a fine-grained WPF attack designed for multi-tab webpage identification from obfuscated traffic. 
Constructing WPF attacks is more challenging than the existing WF attacks because the analyzed webpage traffic patterns exhibit a higher degree of similarity than website traffic patterns.
\ours utilizes metric learning to extract the differences among webpages, which combines proxy-based and sample-based losses to transform the feature space so that samples from the same webpages are clustered and samples from different webpages are separated.
Moreover, we develop data augmentation mechanisms for \ours, which allow \ours to adapt to the diversity of multi-tab traffic in the real world.
We prototype \ours, and evaluate the performance on the collected datasets of multi-tab webpage traffic.
The experiment results demonstrate that \ours achieves 88.6\% and 76.7\% improvements of Recall@5 over the state-of-art attacks under both the closed- and open-world settings, while maintaining a stable performance with various scales of monitored webpages.

%-------------------------------------------------------------------------------
\section*{Acknowledgment}

We thank our anonymous reviewers for their helpful comments and feedback. 
The work is supported in part by NSFC
under Grant 62132011, 62472247, and 62425201.
Qi Li is the corresponding author of this paper.

% The USENIX latex style is old and very tired, which is why
% there's no \textbackslash{}acks command for you to use when
% acknowledging. Sorry.

%-------------------------------------------------------------------------------
% \bibliographystyle{ACM-Reference-Format}
% \bibliography{references}

% \bibliographystyle{plain}
% \bibliography{references}

\bibliographystyle{ACM-Reference-Format}
\balance
\bibliography{references}

\appendix
% \input{section/10-appendix}

%%%%%%%%%%%%%%%%%%%%%%%%%%%%%%%%%%%%%%%%%%%%%%%%%%%%%%%%%%%%%%%%%%%%%%%%%%%%%%%%
\end{document}